\documentclass[fleqn,usenatbib]{mnras}
\usepackage{newtxtext,newtxmath}
\usepackage[T1]{fontenc}

\usepackage{amsmath}
\usepackage{epsfig}

\title[Encounters between anisotropic galaxies]{Orbit decay in encounters between anisotropic spherical galaxies of equal mass}

\author[L. Saleh \& J. E. Barnes]{Lucas Saleh$^1$, Joshua E.~Barnes$^2$
\\
$^1$Department of Physics and Astronomy, University of Hawaii at Manoa; \textsf{lsaleh@hawaii.edu}\\
$^2$Institute for Astronomy, University of Hawaii at Manoa; \textsf{barnes@hawaii.edu}}

\date{Accepted 2023 November 21. Received 2023 November 16; in original form 2023 August 7}

\pubyear{2023}

\begin{document}
\label{firstpage}
\pagerange{\pageref{firstpage}--\pageref{lastpage}}
\maketitle

\begin{abstract}
\noindent We investigate the effect of radial anisotropy on the rate of orbit decay in parabolic encounters of identical spherical galaxies. Our galaxy models have Hernquist density profiles and Osipkov--Merritt velocity distributions. We find that radially anisotropic models merge in as little as half the time of their isotropic counterparts. Anisotropic models are more susceptible to tidal deformation; this accelerates the transfer of orbital angular momentum to internal degrees of freedom. Even during the initial approach, the anisotropic models become more distorted, and arrive at pericentre already having lost substantial amounts of angular momentum. Our results may have implications for estimates of merger rates and persistence of tidal tails.
\end{abstract}

\vspace{0.5in}


\section{INTRODUCTION}
\label{introduction}
Galaxy encounters are remarkably \textit{inelastic}. Two equally-massive galaxies falling together on an interpenetrating parabolic orbit are left on a much smaller and more tightly-bound orbit after their first passage, and typically merge soon after their next approach. This was already clear in early numerical experiments (e.g., \citealt{VV1977}; \citealt{White1978}) and has been affirmed many times since. Orbit decay is driven by the transfer of energy and angular momentum from the relative motion of the galaxies to internal motions within each galaxy \citep[Ch.~8]{BT2008}. 

Early studies indicated that the internal velocity structure of interacting galaxies influences the rate of orbit decay. Working with spherical systems, \cite{White1979} found that merging is faster if galaxies spin in the same direction as they orbit one another, and slower if they spin in a retrograde direction. This arises because particles on direct orbits experience a `broad resonance' \citep{TT1972} with the relative motion of the two galaxies, yielding a stronger tidal response. While the role of rotation has been highlighted in subsequent studies of disc-galaxy mergers, little attention seems to have been paid to other options for velocity structure. \cite{White1979} included non-rotating models with both isotropic and circular velocity distributions, and noted that the latter produced merger remnants with more compact cores, but remarked that `[t]he strength of interaction does not, however, seem to depend on the initial velocity structure.'

Subsequent simulations of encounters between spherical galaxies have explored a wide range of choices for mass ratios, initial orbits, and initial density profiles (e.g., \citealt{V1982,V1983,VM1995,MH1997,FB2001}; \citealt{BKMQ2008}; \citealt{Drakos2019}). These studies have been somewhat overshadowed by simulations of disc galaxy encounters, which have become increasingly realistic as they attempt to reproduce observed interacting systems and model the formation of elliptical galaxies by mergers. Yet even in disc simulations, most of the mass is invested in a roughly spherical dark halo. These haloes are usually set up with isotropic velocity distributions, although some workers (e.g., \citealt{Cox2006}, \citealt{SW1999}) added modest amounts of halo rotation. 

A handful of studies have incorporated radial or tangential anisotropy in simulations of galaxy interactions. \cite{McMillan2007} examined encounters of equal-mass disc galaxies embedded in anisotropic haloes, and reported that initial anisotropy had very little effect on the haloes of merger remnants. When simulating interactions between large galaxies and small satellites, \cite{AB2007} found that satellites orbiting radially aniso\-tropic primaries merged in `somewhat shorter' times, while \cite{Vasiliev2022} noted that radial (tangential) anisotropy caused satellite orbits to become more (less) elongated over time. \cite{Rozier2022} and \cite{Vasiliev2024} included halo anisotropy in studies of the Milky~Way--LMC system. Both found that radially anisotropic halos respond more strongly to the LMC's gravitational field; the latter also reported that the present position and velocity of the LMC are consistent with an initial orbit with higher angular momentum if the dark halo is radially anisotropic.

In this paper we return to the study of encounters between identical spherical systems, and explore the influence of radial anisotropy on orbit decay. We focus on spherical systems for simplicity, and because there are straightforward, accurate ways to generate radially anisotropic models of such systems. Our immediate goal is to clarify the physics of orbit decay, rather than to model real galaxies. However, we note that inasmuch as dark haloes form partly by radial collapse, they are likely to have velocity distributions with preferentially radial orbits.

Our galaxy models, simulation methodology, and parameter choices are described in \S~\ref{sec:methods}. In \S~\ref{sec:results} we present results for simulated encounters and simplified models that explore the response to tidal perturbations. \S~\ref{sec:discussion} presents our conclusions. Appendix~A discusses simulation uncertainties.

\section{METHODS}
\label{sec:methods}

In concept, our experimental approach is very simple. We construct spherical Hernquist (\citeyear{Hernquist1990}) models with anisotropic velocity distributions following Osipkov (\citeyear{Osipkov1979}) and Merritt (\citeyear{Merritt1985}). Two such models are launched on a parabolic collision trajectory, and their evolution is followed until they have merged. We vary the amount of anisotropy while holding other parameters fixed, and use the time between first and second passages to measure the timescale of orbit decay.

To implement this approach, we must make a number of approximations. In particular, (1) the model density profile is exponentially tapered toward zero at a finite radius, (2) dynamical evolution is simulated using an N-body method with finite particle number, (3) gravitational interactions are `softened' at small separations, (4) forces are evaluated using a hierarchical tree code which trades accuracy for speed, (5) particle trajectories are integrated using a finite time-step, and (6) the models are launched from a finite separation. We discuss each of these approximations below and estimate their effects on our results.

\subsection{Mass Distribution and Gravitational Potential}
\label{sec:mass_model}

We use the Hernquist (\citeyear{Hernquist1990}) density profile to model a simple, spherically symmetric galaxy. This profile is fully characterized by two parameters, the mass $M$ and scale radius $a$. The density is 
\begin{equation}
    \rho_\mathrm{H}(r)=\frac{1}{2 \pi} \, \frac{M a}{r (r+a)^3} \, .
    \label{eq:hernquist_profile}
\end{equation}
The corresponding gravitational potential is 
\begin{equation}
    \Phi_\mathrm{H}(r)= - \, \frac{G M}{r+a} \, .
    \label{eq:hernquist_potential}
\end{equation}
where $G$ is the gravitational constant. 

For simplicity, we use units in which $G = M = a = 1$. This entails no loss of generality, as the resulting models can be rescaled to any desired system of units. In these units, the half-mass radius is $r_\mathrm{h} = 1 + \sqrt{2}$. A circular orbit at this radius has period $P_\mathrm{h} \simeq 33.332$; this is a characteristic timescale for dynamical evolution.

At large radii, $\rho_\mathrm{H}(r)$ falls off as $\rho \propto r^{-4}$. When this density profile is sampled with $N$ particles, the outermost one will have radius $r \sim a N$. Such extreme outliers are numerically inconvenient, so we use a general tapering function \citep[eq.~16]{Barnes2012} to smoothly taper the density profile, starting at a radius $b = 100 a$. This effectively limits the outermost particle to a radius of roughly $500 a$. To keep the total mass of the model fixed, matter at large radii is redistributed within $r < b$; this increases the density within radius $b$ by just under $1$ per~cent. In what follows, $\rho(r)$ is the tapered density profile.

Gravitational interactions between particles are softened at short range using the usual Plummer formula: at a distance $R$, the potential of a particle is proportional to $(R^2 + \epsilon^2)^{-1/2}$, where $\epsilon$ is a small `softening length'. This biases the overall gravitational potential, so $\Phi_\mathrm{H}(r)$ is no longer accurate. Since we need accurate potentials to compute velocity distributions (\S~\ref{sec:velocity_distribution}), the tapered density profile $\rho(r)$ is numerically convolved with a Plummer kernel \citep[eq.~6]{Barnes2012}, and the result is used to integrate the potential as a function of radius. In what follows, $\Phi(r)$ is the resulting potential.

\subsection{Velocity Distribution}
\label{sec:velocity_distribution}

We use the Osipkov--Merritt method to construct radially anisotropic DFs (distribution functions) for our models \citep{Osipkov1979,Merritt1985}. For a spherical isotropic system, the DF is $f(\mathbf{r}, \mathbf{v}) = f(E)$, where $E \equiv \frac{1}{2} |\mathbf{v}|^2 + \Phi(|\mathbf{r}|)$ is the specific binding energy. The Osipkov--Merritt DF is $f_\mathrm{a}(\mathbf{r}, \mathbf{v}) = f_\mathrm{a}(Q)$, where $Q \equiv E + \frac{1}{2} J^2 / r_a^2$; here, $J$ is the specific angular momentum and $r_\mathrm{a}$ is the anisotropy radius, a parameter which controls the degree of anisotropy. Resolving the velocity into a radial component $v_\mathrm{r}$ and a tangential component $v_\mathrm{t}$, we obtain
\begin{equation}
    Q = \frac{1}{2} v_\mathrm{r}^2 +
        \frac{1}{2} \left(1 + \frac{r^2}{r_\mathrm{a}^2} \right) v_\mathrm{t}^2 + \Phi(|\mathbf{r}|) \, .
\end{equation}
The velocity distribution at any radius is stratified on prolate spheroids. For $r \ll r_\mathrm{a}$, the quantity $Q \simeq E$ and the velocity distribution is approximately isotropic. Conversely, for $r \gg r_\mathrm{a}$, transverse velocities are weighted more than radial velocities, so the velocity distribution is preferentially radial. The anisotropy at radius $r$ is quantified by
\begin{equation}
    \beta(r) \equiv 1 - \frac{\langle v_t^2 \rangle}{\langle 2 v_r^2 \rangle}
             = \frac{r^2}{r^2 + r_\mathrm{a}^2} \, ,
\end{equation}
where the final equality is specific to Osipkov--Merritt DFs. Isotropic distributions have $\beta = 0$, while radially anisotropic distributions have $0 < \beta \le 1$.

To compute the DF, given a density $\rho(r)$, potential $\Phi(r)$, and anisotropy radius $r_\mathrm{a}$, we first set
\begin{equation}
    \tilde{\rho}(r) = \left( 1 + \frac{r^2}{r_\mathrm{a}^2} \right) \rho(r) \, .
\end{equation}
Since the potential is a monotonic function of radius, $\tilde{\rho}$ can be considered a function of $\Phi$. The DF is then given by
\begin{equation}
    f_\mathrm{a}(Q) = \frac{1}{\sqrt{8} \pi^2} \, \frac{\mathrm{d}}{\mathrm{d}Q}
        \int_Q^0 \frac{1}{\sqrt{\Phi - Q}} \,
            \frac{\mathrm{d}\tilde{\rho}}{\mathrm{d}\Phi} \, \mathrm{d}\Phi \, .
            \label{eq:OM_formula}
\end{equation}

In practice, these steps must be performed numerically, since the potential $\Phi(r)$ is not available analytically. We use spline interpolation to invert $\Phi(r)$ and compute $\mathrm{d}\tilde{\rho} / \mathrm{d}\Phi$. An adaptive routine is used to compute the integral in equation~(\ref{eq:OM_formula}); the result is tabulated as a function of $Q$, and spline interpolation is used for the final differentiation. When given $\rho_\mathrm{H}(r)$ and $\Phi_\mathrm{H}(r)$ as inputs, our numerical procedure reproduces the corresponding Osipkov--Merritt DF \citep{Hernquist1990} with relative errors $< 0.1$ per~cent.

\subsection{N-body Simulations}
\label{sec:nbody}

We use self-consistent N-body simulations to follow the interactions and mergers of our anisotropic models. This approach represents the DF as a collection of $N$ particles; particle $i = 1, \dots, N$ has mass $m_i = M / N$, position $\mathbf{r}_i$, and velocity $\mathbf{v}_i$. Each particle obeys Newton's laws of motion in the gravitational field generated by all other particles.

Particles are initialized so that the probability of finding one with coordinates $(\mathbf{r}_i, \mathbf{v}_i)$ is proportional to the value of $f_\mathrm{a}(\mathbf{r}_i, \mathbf{v}_i)$. Each particle is initialized independently as follows. First, the particle's radius $r_i$ is chosen by generating a random number $M_i$ uniformly distributed in the range $[0, M)$, and then solving $M_i = M(r_i)$, where $M(r') = \int_{0}^{r'} 4 \pi r^2 \rho(r) dr$ is the cumulative mass profile. We then generate a random unit vector $\hat{\mathbf{u}}_i$ and set the particle position $\mathbf{r}_i = r_i \, \hat{\mathbf{u}}_i$. Second, at radius $r_i$, we use rejection sampling to pick a speed $v'_i$ from the speed distribution $v^2 f_\mathrm{a}(\Phi(r_i) + \frac{1}{2} v^2)$. Another random unit vector $\hat{\mathbf{u}}'_i$ is generated, and we set $\mathbf{v}'_i = v'_i \, \hat{\mathbf{u}}'_i$. Finally, the tangential components of $\mathbf{v}'_i$ are rescaled by a factor of $(1 + r_i^2 / r_\mathrm{a}^2)^{-1/2}$ to yield the particle velocity $\mathbf{v}_i$.

The first and second steps of this procedure, which suffice to generate isotropic N-body realizations, were extensively tested by \citet{Barnes2012}. The final step, which transforms an isotropic (spherical) velocity distribution into the spheroidal distribution required by the Osipkov--Merritt formalism, is new. Stability tests, to be described in the next section, confirm that our initial conditions are very close to equilibrium. 

In our simulations we use $N = 65536 = 2^{16}$ particles for each spherical model. Power-of-two values for $N$ ensure that the particle mass $m_i$, and integer multiples thereof, are exactly represented as floating point numbers. Our specific choice for $N$ represents a compromise between sampling accuracy and computing resources. Since particles are initialized independently, integrated quantities will have Monte-Carlo uncertainties of order $N^{-1/2}$, or $\sim 0.4$ per~cent for our value of $N$. `Quiet-start' procedures which suppress Monte-Carlo fluctuations are available for disc simulations \citep[e.g.,][]{DS2000}, but not for a general $3$-D problem like the encounters we study here. We therefore ran multiple realizations of each experiment with different random seeds to initialize the particle coordinates, and used run-to-run variation to estimate the uncertainty due to sampling.

Gravitational forces are computed using a tree algorithm \citep{BH1986}. In this algorithm, the force on each particle is approximated by a sum of $O(\log N)$ terms, which represent nearby particles and a hierarchy of cubical cells at larger distances. To be included in this hierarchy, a cell at a distance $R$ from a given particle must satisfy $d < \theta R$, where $\theta$ is a fixed parameter which controls the trade-off between speed and accuracy. We set $\theta = 0.75$ and expand the gravitational field of each cell to quadrupole order. With these choices, the median acceleration error is $\sim 0.03$ per~cent.

As noted above, we adopt Plummer softening to limit gravitational accelerations due to short-range interactions. We use a softening length of $\epsilon = 1/64$, which is small enough to resolve the inner $\rho \propto r^{-1}$ slope of the Hernquist model \citep{Barnes2012}. Compared to the original potential (equation~\ref{eq:hernquist_potential}), this softening biases the central potential upward by $\sim 3$ per~cent. However, at radius $r = a$, the potential bias is only $\sim 0.1$ per~cent, and at larger radii it's even smaller. Since the models in our simulated encounters are typically separated by distances greater that $a$, softening is unlikely to have any significant impact on our results.

We use a time-centred leapfrog with a uniform time-step to compute particle trajectories. This simple scheme is adequate for our purposes since softening limits the rate of change of particle accelerations. Our time-step is $\delta t = 1/64 = 2^{-6}$ (again, a power-of-two value guarantees that times have exact representations). With this $\delta t$, our simulations conserve energy and angular momentum to better than $0.1$ per~cent.

All our calculations were run on $4$~GHz Intel processors, using single-precision arithmetic. A typical encounter simulation, following $131072$ particles for $500$ time units, took $\sim 58$~hours of CPU time.

\subsection{Stability}
\label{sec:stability}

Spherical systems with predominantly radial orbits are often unstable, spontaneously evolving into roughly prolate configurations (\citealt{Antonov1973}; \citealt{MA1985}; \citealt{BGH1986}). While it's certainly possible to simulate encounters of unstable systems, it's not easy to motivate such experiments as analogs of encounters between real galaxies; to the extent that galaxies can be described as equilibrium systems, we expect them be stable or nearly so. Thus our initial simulations were designed to determine the maximum amount of anisotropy our models can bear without becoming noticeably unstable. 

In addition to visually inspecting individual simulations, we quantified the effects of the radial orbit instability as follows. After generating each model realization (\S~\ref{sec:nbody}), all $N$ particles are sorted by binding energy $E$; this ordering will be largely preserved as the system evolves. Sectioning the particle array into $8$ equal segments thus defines $8$ Lagrangian volumes or `shells' ordered by binding energy. The shape of shell $j$ is measured by computing the minor-to-major axis ratio $(c/a)_{j}$ from eigenvalues of the shell's moment of inertia tensor. The most tightly-bound shells are relatively poor indicators of instability, remaining nearly spherical even when more outlying shells are clearly distorted. We therefore used the $7^\mathrm{th}$ shell (i.e., the $75^\mathrm{th}$ to $87.5^\mathrm{th}$ percentile), which primarily samples the mass distribution at radii between $\sim 5 a$ and $\sim 14 a$, to track the development of the radial-orbit instability. (The $8^\mathrm{th}$ shell was not useful, as it includes a few very distant particles which individually bias the moment of inertia tensor; these particles have orbital periods \textit{much} longer than the duration of our simulations.) 

A trial run with $r_\mathrm{a} = 1$ indicated that this model is unstable. We therefore tested anisotropy radii $r_\mathrm{a} = 2^{k/8}$ where $k > 0$ is an integer. This scheme samples $r_\mathrm{a}$ finely for small $k$ and more coarsely for larger $k$. Single realizations with $k = 1, \dots, 8$ were run using the N-body parameters given in \S~\ref{sec:nbody}. Models with $k \ge 4$ (i.e., $r_\mathrm{a} \ge 1.414$) appeared stable, with $(c/a)_{7}$ values fluctuating between $\sim 0.95$ and unity. In contrast, models with $k \le 2$ (i.e., $r_\mathrm{a} \le 1.189$) clearly evolved towards more non-spherical shapes, typically falling below $(c/a)_{7} < 0.85$ after $500$ time units ($\sim 15$ orbit periods at the half-mass radius).

\begin{figure}
    \centering
    \includegraphics[width=\columnwidth]{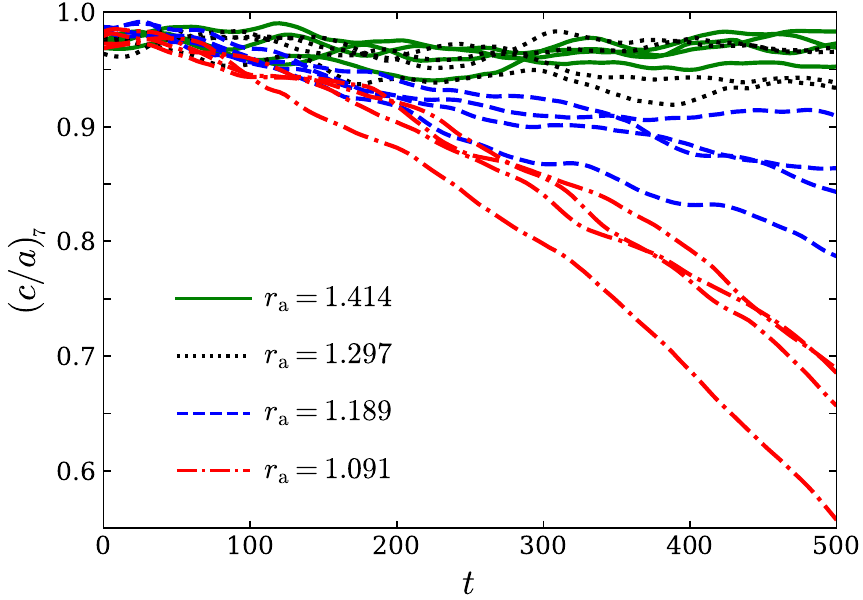}
    \caption{Evolution of $(c/a)_{7}$, the minor-to-major axial ratio of shell $7$, for models with various anisotropy radii $r_\mathrm{a}$. Four independent realizations are shown for each choice of $r_\mathrm{a}$.}
    \label{fig01}
\end{figure}

To better constrain the boundary of stability, we ran another three realizations of each model with $k = 1, \dots, 4$. Fig.~\ref{fig01} presents the results. The additional runs confirm that models with $k \le 2$ are unstable, while those with $k \ge 4$ appear to be stable. As a group, the models with $k = 3$ (i.e., $r_\mathrm{a} = 1.297$) typically maintain axial ratios $(c/a)_{7} \simeq 0.95$. Individual realizations fluctuate about this value, but none of the four realizations displays the systematically decreasing $(c/a)_{7}$ ratios characteristic of the unstable models. 

After completing our tests, we learned that \cite{MZ1997} and \cite{Buyle2007} also studied the stability of anisotropic Hernquist models with Osipkov--Merrit DFs. \citeauthor{MZ1997} estimated the stability boundary at $r_\mathrm{a} \simeq 1.0$, while \citeauthor{Buyle2007} placed it at $r_\mathrm{a} \simeq 1.1$. Our results indicate that both of these models are unstable. The reason for this discrepancy may lie in the method used to detect the instability. Both of the earlier studies quantified the shapes of their simulations using the inner $\sim 70$ per~cent of the mass distribution. Measuring shapes at smaller radii may have led these studies to overlook weak instabilities, which manifest most strongly at large radii.

We adopt the $r_\mathrm{a} = 1.297$ model as the most anisotropic which is dynamically stable enough to be useful in this work. We cannot preclude the possibility that this model is unstable on timescales longer than the duration of our experiments; nor can we preclude an instability which manifests only at extremely large radii. Rather, we argue that such weak instabilities would have little practical consequence for our experiments. In support of this, we note that the merging behavior of our `maximally anisotropic' model is continuous with the behavior of less anisotropic models.

\subsubsection{Initial equilibria}
\label{sec:initial_equilibria}

As mentioned above, the stability calculations can also be analysed to check our procedure for initializing particle coordinates. Errors in generating equilibrium initial conditions generally manifest on a dynamical timescale. We therefore looked for abrupt evolution in scatter-plots of particle radii $r$ vs.~radial velocity $v_r$ or tangential velocity $v_t$, in the ratio of velocity dispersions $\langle v_t^2 \rangle / \langle 2 v_r^2 \rangle$ for particles in nested spherical shells, and in radial mass profiles. No changes on a dynamical timescale were observed, implying our initial conditions, whether stable or not, are very close to equilibrium. In stable anisotropic models, we did notice modest evolution of the ratio $\langle v_t^2 \rangle / \langle 2 v_r^2 \rangle$, consistent with two-body relaxation slowly driving this quantity toward its isotropic value of unity. However, the duration of our simulations is too short to accurately characterize this effect.

\subsection{Encounters}
\label{sec:Encounters}

\newcommand{\texthalf}{\textstyle\frac{1}{2}}

Our experiments mimic encounters of galaxies falling together on asymptotically parabolic trajectories. Initial positions and velocities are obtained from an idealized parabolic orbit, with pericentric separation $r_\mathrm{p}$, of two point-like bodies each of mass $M = 1$. We adopt a coordinate system in which these bodies move clockwise in the $(x,y)$ plane. At pericentre, which defines time $t = 0$, they are symmetrically placed on the $x$-axis, with positions $x = \pm \texthalf r_\mathrm{p}$ and velocities $v_y = \mp \sqrt{G M / r_\mathrm{p}}$. We then track the bodies back to a chosen starting time $t_{0} < 0$, yielding their initial separation vector $\mathbf{r}_{0}$ and relative velocity vector $\mathbf{v}_{0}$.

We next replace these point-like bodies with N-body realizations of the anisotropic DF $f_\mathrm{a}(\mathbf{r}, \mathbf{v})$ for a given anisotropy radius $r_\mathrm{a}$. A different random number seed is used for each realization. These realizations are then translated by $\pm \frac{1}{2} \mathbf{r}_{0}$ in position, and $\pm \frac{1}{2} \mathbf{v}_{0}$ in velocity, and superimposed. Thus, the initial conditions realize a system with the DF
\begin{equation}
    f_\mathrm{sys} =
      f_\mathrm{a}(\mathbf{r} - \texthalf \mathbf{r}_{0},
        \mathbf{v} - \texthalf \mathbf{v}_{0}) +
      f_\mathrm{a}(\mathbf{r} + \texthalf \mathbf{r}_{0},
        \mathbf{v} + \texthalf \mathbf{v}_{0}) \, .
    \label{eq:sys_dist_func}
\end{equation}
Note that in constructing our initial conditions, we have \textit{not} subtracted centre-of-mass positions and velocities from the individual galaxy realizations. As noted in \S~\ref{sec:nbody}, bulk quantities have uncertainties due to Monte-Carlo statistics; as a result, each galaxy realization has an initial offset in position and velocity, of order $O(N^{-1/2})$, with respect to its assigned coordinates. While these offsets are small, they do have measurable consequences, as we show in Appendix~A.

\begin{table}
    \centering
    \caption{Encounter simulations.  Primary parameters are the pericentric separation of the initial orbit, $r_\mathrm{p}$, and the Osipkov--Merritt anisotropy radius, $r_\mathrm{a}$ (where $r_\mathrm{a} = \infty$ indicates an isotropic model). The starting time is $t_{0}$. We ran $N_\mathrm{real}$ realizations of each encounter.}
    \begin{tabular}{l l l c}
        \noalign{\bigskip} \hline \noalign{\smallskip}
        $r_\mathrm{p}$
        &
        $r_\mathrm{a}$
        &
        $t_{0}$
        &
        $N_\mathrm{real}$ \\
        \noalign{\smallskip}
        \hline
        \noalign{\smallskip}
        1.0 & 1.297    & -200 & 8 \\
            & 1.297    & -400 & 8 \\
            & 1.414    & -200 & 8 \\
            & 2.0      & -200 & 4 \\
            & 4.0      & -200 & 4 \\
            & 8.0      & -200 & 4 \\
            & $\infty$ & -200 & 8 \\
            & $\infty$ & -400 & 8 \\
        \noalign{\medskip}
        2.0 & 1.297    & -200 & 8 \\
            & 1.297 & -400 & 8 \\
            & 1.414    & -200 & 4 \\
            & 2.0      & -200 & 4 \\
            & 4.0      & -200 & 4 \\
            & 8.0      & -200 & 4 \\
            & $\infty$ & -200 & 8 \\
            & $\infty$ & -400 & 8 \\
        \noalign{\medskip}
        4.0 & 1.297    & -200 & 8 \\
            & 1.297    & -400 & 8 \\
            & 1.414    & -200 & 4 \\
            & 2.0      & -200 & 4 \\
            & 4.0      & -200 & 4 \\
            & 8.0      & -200 & 4 \\
            & $\infty$ & -200 & 8 \\
            & $\infty$ & -400 & 8 \\
            \hline
    \end{tabular}
    \label{tab:simulations}
\end{table}

The choice of starting time requires some attention. To be entirely consistent with the chosen two-body orbit, the galaxies should ideally start so far apart that they initially interact like disjoint spherical systems. This isn't possible for models with extended outer envelopes; some overlap has to be tolerated. We quantified the degree of overlap using
\begin{equation}
    \delta U = U_\mathrm{tot} - (U_{1} + U_{2} - G M^2 / r_{0}) \, ,
    \label{eq:deltaU}
\end{equation}
where $U_\mathrm{tot}$ is the total potential energy of the initial configuration, and $U_{1}$ and $U_{2}$ are the internal potential energies of the two models. Separate spherical systems yield $\delta U = 0$; overlap correlates with $\delta U > 0$.

Our experiments start $t_{0} = -200$ time units before pericentre, unless another time is explicitly stated. This places the models $r_{0} \simeq 69$ length units apart. For this configuration, the resulting $\delta U$ was $\sim 1.3$ per~cent of $G M^2 / r_{0}$, which seemed a bit large but acceptable inasmuch as the same bias applies to all models with the same $t_{0}$ and $r_\mathrm{p}$. To check, we re-ran some key experiments starting at $t_{0} = -400$. This initially placed the models $\sim 111$ length units apart; the resulting $\delta U$ was $\sim 0.42$ per~cent of $G M^2 / r_{0}$. 

Table~\ref{tab:simulations} lists parameters for our encounter simulations. We ran at least four realizations of each encounter, and an additional four for a subset. The extra realizations were run to test both choices for $t_{0}$ at the extremes for $r_\mathrm{p}$ and $r_\mathrm{a}$, and to anchor our results by accurately characterizing the most extreme cases.

\subsubsection{Orbital trajectories}
\label{orbital_trajectories}

To follow the process of orbit decay, we extract trajectories from the encounter simulations. Our method relies on the fact that particles which are initially tightly bound remain tightly bound as the system evolves. As noted in \S~\ref{sec:stability}, the particles in each realization are initially sorted in binding energy $E$, with the most tightly bound particles first. At each time-step, we can therefore estimate the position and velocity of each realization by averaging over the coordinates of its first $N_\mathrm{c}$ particles. Our choice of $N_\mathrm{c} \, = 4096$ represents a trade-off between statistical and systematic uncertainties; larger values would yield smoother trajectories, but loosely-bound particles could be detached from their centres following a close passage, potentially biasing the measured coordinates at later times. Mean and median positions for representative samples of $N_\mathrm{c} = 4096$ tightly-bound particles, taken from a large ensemble of simulations, typically agree to $\sim 0.005$ length units ($\sim 30$~per cent of the softening length, $\epsilon$), indicating that detached outliers have little effect. Key results are also robust against halving or doubling $N_\mathrm{c}$.

\begin{figure}
    \centering
    \includegraphics[width=60mm]{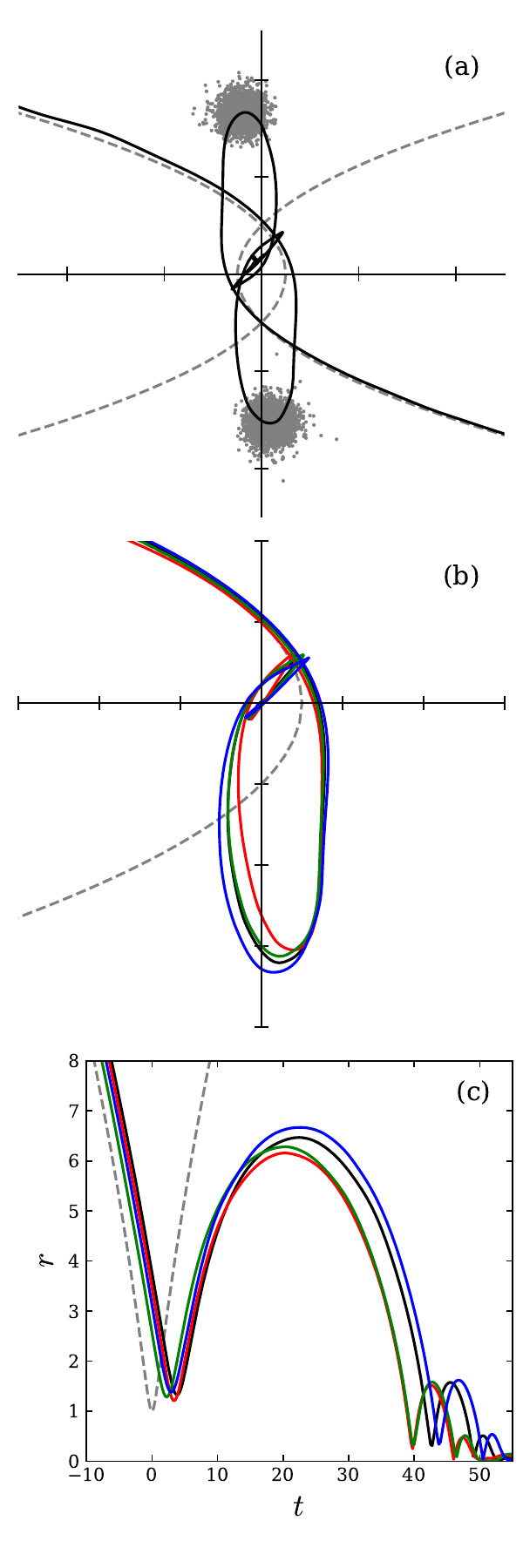}
    \caption{Trajectories from encounters of isotropic models with $r_\mathrm{p} = 1$.  In all panels, dashed curves show idealized parabolic orbits. Panel (a): a single simulation. Grey dots show positions for a tightly-bound sample of particles from each model, plotted $20$ time units after pericentre. Solid black curves are trajectories derived from these samples. Panel (b): relative trajectories of four equivalent simulations, illustrating run-to-run variations. Panel (c): radial trajectories (separation $r$ vs.~time $t$) for the runs shown in~(b).}
    \label{fig02}
\end{figure}

Panel~(a) of Fig.~\ref{fig02} shows how this works, using an encounter of two isotropic models with $r_\mathrm{p} = 1$. Here, the grey points are the most tightly bound $N_\mathrm{c}$ particles in each realization, shown $t = 20$ time units after pericentre. These particle distributions are still symmetric and well-localized about the centres of their respective realizations. The solid curves are the trajectories extracted by averaging the positions of these particles at each time-step, while the dashed curves are the idealized parabolic orbits used to set up the initial conditions. All of our encounter simulations unfold in a broadly similar fashion, with the two models approximating idealized parabolic trajectories up to first passage, reaching apocentre shortly thereafter, and subsequently falling back for a second and much closer passage.

Panel~(b) shows four simulations, all sampling the same initial DF as in panel~(a), but initialized using different random number seeds. Here we plot a single relative trajectory for each simulation, fixing the other model at the origin; this reduces visual clutter at the cost of suppressing the slight asymmetries between individual trajectories seen in the top panel. The variation from run-to-run is due to Monte-Carlo sampling of the initial conditions; to narrow the range, we would need to increase the number of particles, $N$.

Panel~(c) plots separations as functions of time for the same ensemble of simulations. We note that the actual time of first pericentre (hereafter $t_\mathrm{p1}$) is delayed relative to the pericentre of the idealized parabolic trajectory, which occurs at $t_\mathrm{p} = 0$. Moreover, the actual pericentric separation (hereafter $r_\mathrm{p1}$) is larger than the idealized separation of $r_\mathrm{p} = 1$. Conservation of orbital energy and and angular momentum predicts differences of this kind for an interpenetrating encounter of two rigid spheres. As we will see, the actual dynamics are more complex; tides deform the models even during their initial approach, transferring energy and angular momentum to internal degrees of freedom. 

Panel~(c) also shows that each encounter has a well-defined second pericentre at time $t_\mathrm{p2}$, and merges shortly thereafter. We will use $\Delta t = t_\mathrm{p2} - t_\mathrm{p1}$ to measure the timescale for orbit decay.

\section{RESULTS}
\label{sec:results}

\begin{figure*}
    \centering
    \includegraphics[width=\textwidth]{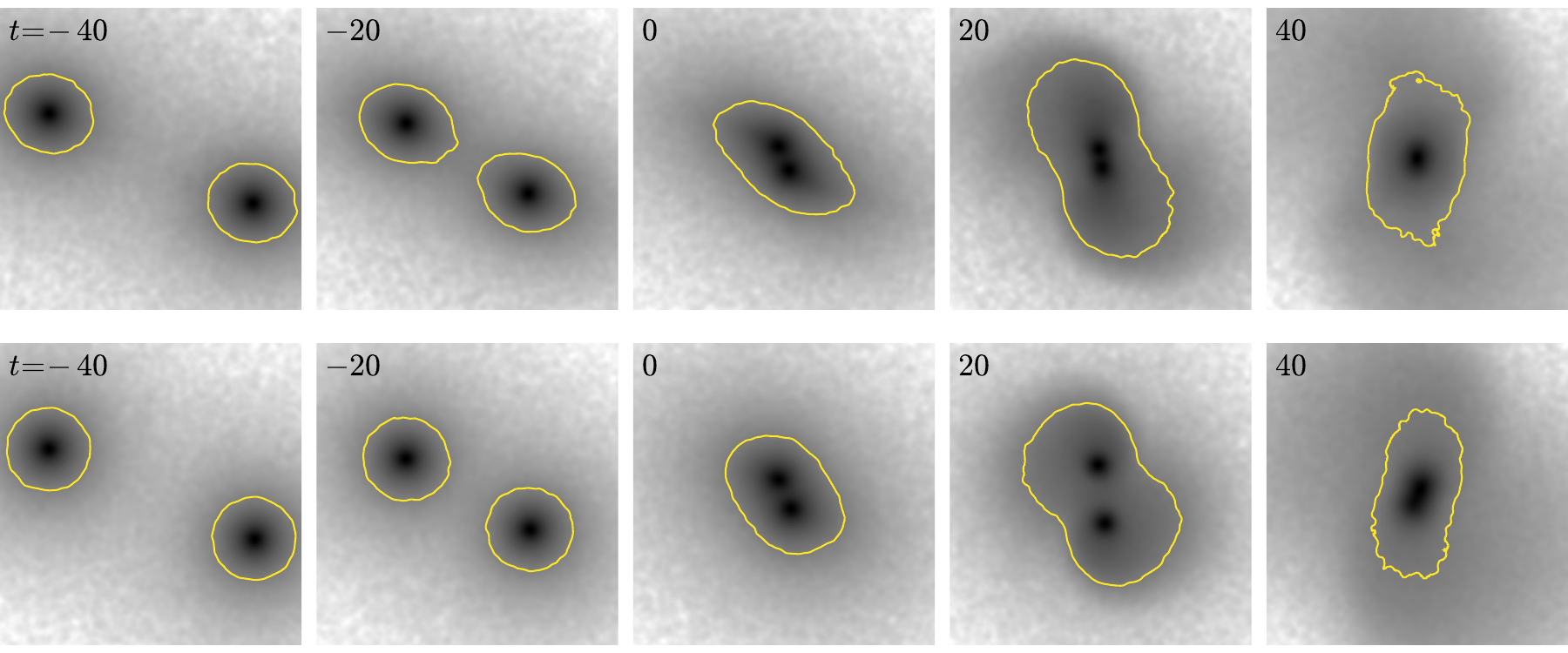}
    \caption{Encounters between anisotropic models (top; $r_\mathrm{a} = 1.297$) and isotropic models (bottom). In both cases, the galaxies were launched on parabolic orbits with $r_\mathrm{p} = 1$. Each image is $32$ length units on a side, and shows a stack of eight realizations. A single contour is included to show the tidal distortion. The anisotropic models fall back together much more rapidly, as the later panels show.}
    \label{fig03}
\end{figure*}

Fig.~\ref{fig03} compares encounters of anisotropic models (top row; $r_\mathrm{a} = 1.297$) with equivalent encounters of isotropic models (bottom row). These models were all launched with the \textit{same} initial positions and velocities, taken from a parabolic encounter of two point masses with pericentric separation $r_\mathrm{p} = 1$. Each frame is a stack of eight independent realizations, with a single contour of projected particle density tracing the shapes of the participants. This figure demonstrates that \textit{anisotropy has a significant effect on orbit decay.}

The frames on the left depict the models during their approach. Some effects of anisotropy are already visible at these early stages, especially at time $t = -20$, where the contour shows that the anisotropic models are already tidally distorted, while their isotropic counterparts remain nearly spherical. The middle frames, at $t = 0$, show both pairs of models deeply interpenetrating just before first passage; while these images are superficially similar, the anisotropic models continue to show marked distortion along the axis of their initial approach. After first passage, the effects of anisotropy become more dramatic: at $t = 20$, the isotropic models are loitering near apocentre, while the anisotropic models are well past apocentre and plunging toward their \textit{second} passage. In the final frames, at $t = 40$, the anisotropic models have already merged, while the isotropic models are just beginning their second encounter.

\subsection{Orbital Evolution}
\label{sec:orbital_evolution}

\begin{figure*}
    \centering
    \includegraphics[width=\textwidth]{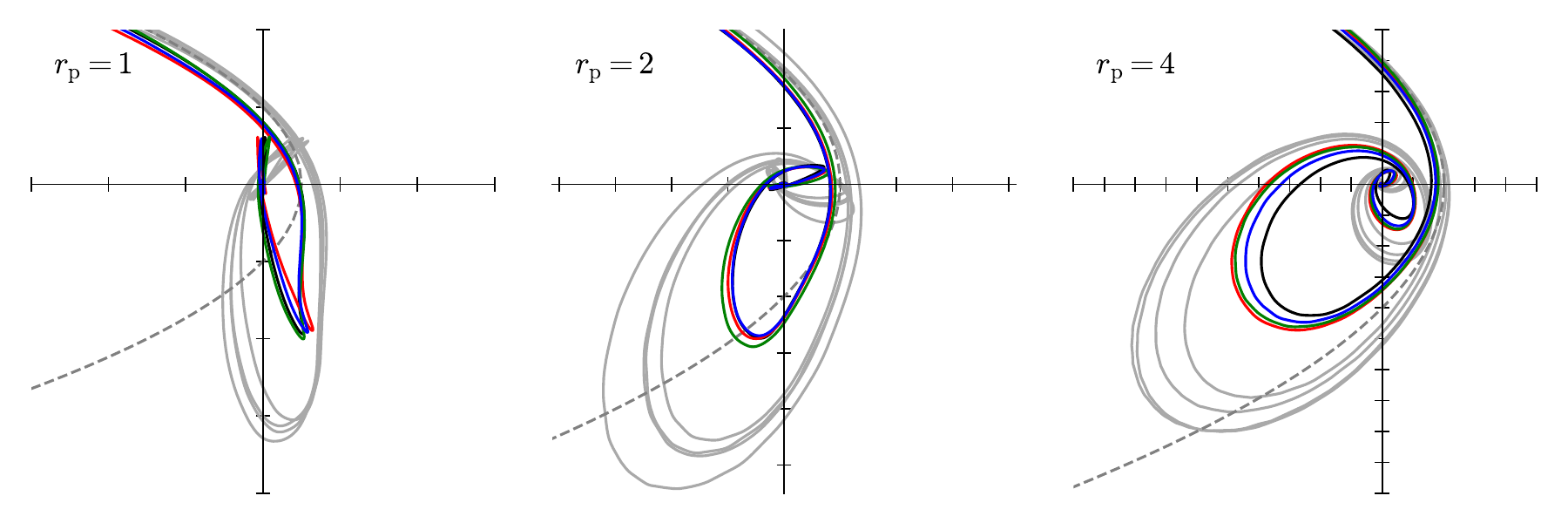}
    \caption{Trajectories from encounters between anisotropic models (red, green, blue, and black curves; $r_\mathrm{a} = 1.297$) and isotropic models (light grey curves), with pericentric separation $r_p$ increasing from left to right. Four realizations of each encounter are shown. Dashed curves plot idealized parabolic orbits. Tick marks are 2 length units apart. In every case, the anisotropic models reach significantly smaller separations after pericentre.}
    \label{fig04}
\end{figure*}

\begin{figure*}
    \centering
    \includegraphics[width=\textwidth]{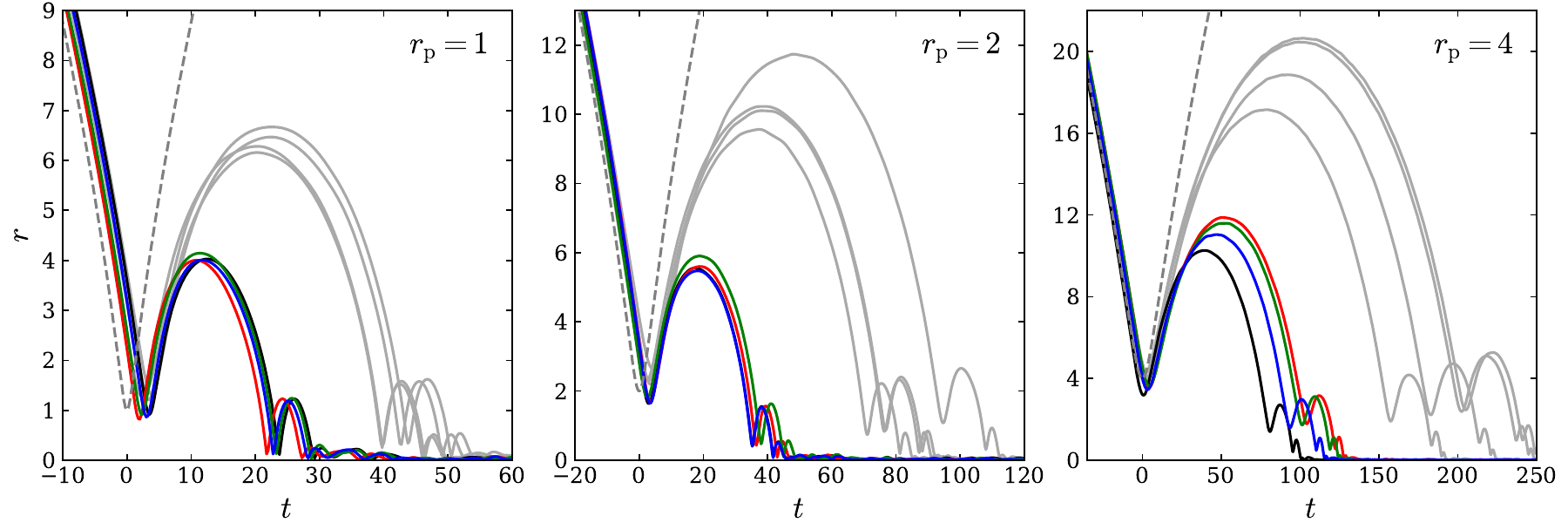}
    \caption{Separation vs.~time for anisotropic and isotropic models. This figure uses the same simulations and color scheme used in Fig.~\ref{fig04}. In every case the anisotropic encounters merge more quickly than their isotropic counterparts. }
    \label{fig05}
\end{figure*}

Fig.~\ref{fig04} compares trajectories from encounters between anisotropic models with their isotropic counterparts. Here, the colored curves represent models with maximal anisotropy $r_\mathrm{a} = 1.297$, while the light grey curves represent isotropic models. Each panel shows results for a different choice of idealized pericentric separation $r_\mathrm{p}$. Across the entire set of encounters, it's evident that anisotropy has a strong effect on orbit decay; this effect is much larger than the run-to-run variation between models in each ensemble. The anisotropic models all attain much smaller separations after first passage, and their subsequent trajectories are more nearly radial than those of their isotropic counterparts. Indeed, the anisotropic encounters with $r_\mathrm{p} = 1$ come to a near-standstill at apocentre, and fall back together on nearly linear trajectories. This is an example of `radialization' -- the tendency for orbits to become \textit{more} eccentric as they decay. Radialization in unequal-mass encounters is discussed by \citet{Amorisco2017} and \citet{Vasiliev2022}; the latter study showed that radial anisotropy in the host \textit{accelerates} the radialization of satellite orbits. \citet{Barnes2016} illustrates an extreme example of radialization, in which two equal-mass galaxies with isotropic halos \textit{reverse} their sense of orbital motion after first passage; passages closer than those considered here would likely exhibit this reversal as well.

Careful inspection of Fig.~\ref{fig04} reveals a curious trend; during the initial approach, the trajectories of the anisotropic models lie inside the idealized parabolic orbits, while those of the isotropic models generally fall on the outside. This is not the result of run-to-run variations or a problem with the initial conditions; \textit{orbit decay is modifying the trajectories of these models even before first passage}. 

Fig.~\ref{fig05} shows the same simulations, with separations plotted as functions of time. This figure again shows that the anisotropic simulations have much smaller orbits after the first passage. As one might expect, the encounters between highly anisotropic systems also reach apocentre and fall back to a second pericentre much faster than their isotropic counterparts. 

\begin{figure*}
    \centering
    \includegraphics[width=\textwidth]{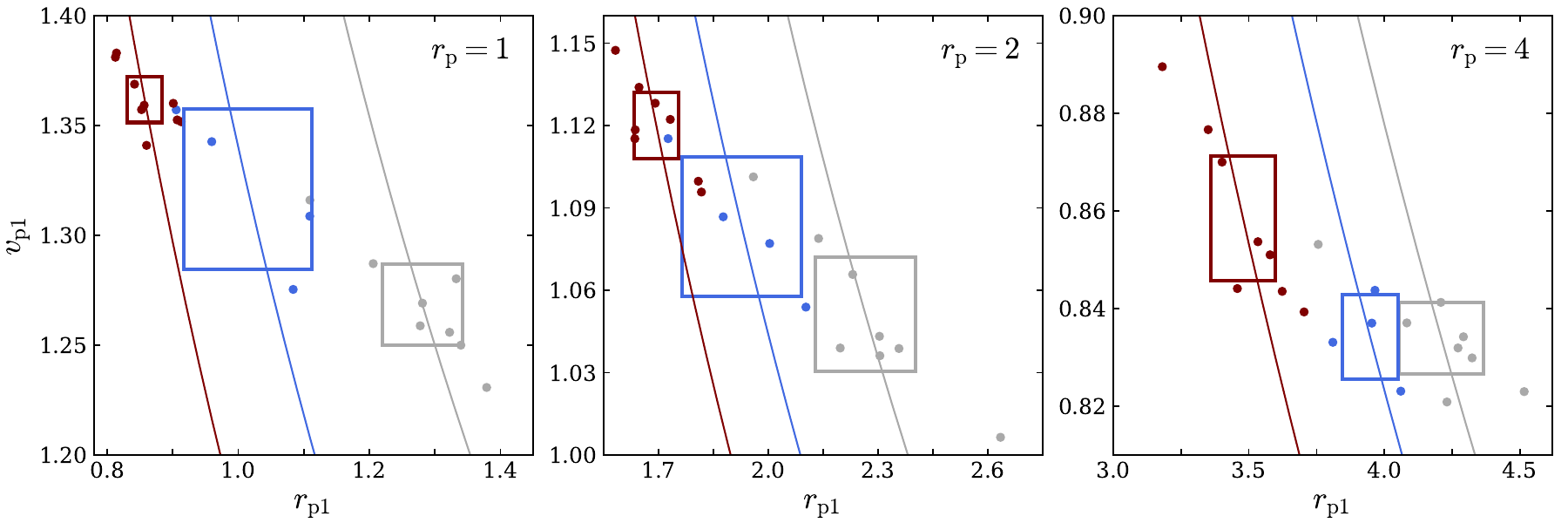}
    \caption{Dynamical parameters at first passage. Here, $r_\mathrm{p1}$ is the minimum separation of the galaxies, while $v_\mathrm{p1}$ is their relative velocity at that time. Red, blue, and grey represent models with $r_\mathrm{a}=1.297$, $2.0$, and $\infty$, respectively. Points show individual simulations, while boxes show sample means with $2 \sigma$ uncertainties. Curves are contours of constant orbital angular momentum. Reducing $r_\mathrm{a}$ decreases pericentric separation and increases relative velocity.}
   \label{fig06}
\end{figure*}

We focus on first pericentre passages in Fig.~\ref{fig06}, which plots relative velocities $v_\mathrm{p1}$ vs.~separations $r_\mathrm{p1}$ for all three values of $r_\mathrm{p}$. This figure shows results for isotropic models (black; $r_\mathrm{a} = \infty$), moderately anisotropic models (blue; $r_\mathrm{a} = 2$), and maximally anisotropic models (red; $r_\mathrm{a} = 1.297$). Between $N_\mathrm{real} = 4$ and $8$ realizations of each encounter are plotted as points to show run-to-run variations. For each ensemble, we plot a box centered on the sample means $\overline{r_\mathrm{p1}}$ and $\overline{v_\mathrm{p1}}$. Each box extends $\pm 2 N_\mathrm{real}^{-1/2} s_{r}$ in the horizontal direction, and $\pm 2 N_\mathrm{real}^{-1/2} s_{v}$ in the vertical direction, where $s_{r}$ and $s_{v}$ are sample standard deviations in $r_\mathrm{p1}$ and $v_\mathrm{p1}$, respectively (estimated assuming $1$ degree of freedom). Thus, to the extent that the central limit theorem applies for small samples, the boxes indicate $\sim 2 \sigma$ uncertainties in the sample means. It is evident that pericentric separation and velocity are both influenced by anisotropy, with smaller $r_\mathrm{a}$ values yielding closer and faster passages.

The systematic trends with $r_\mathrm{a}$ seen here imply that a significant amount of orbital evolution is taking place during the approach to first passage. For equal-mass encounters, the specific orbital angular momentum at first passage is $J_\mathrm{p1} = \texthalf r_\mathrm{p1} v_\mathrm{p1}$; the curves plotted in Fig.~\ref{fig06} are contours of constant angular momentum, passing through the mean $\overline{r_\mathrm{p1}}$ and $\overline{v_\mathrm{p1}}$ of each sample. Anisotropic models clearly lose more orbital angular momentum than their isotropic counterparts. This difference is most dramatic for close passages; among the $r_\mathrm{p} = 1$ encounters, the most anisotropic models have lost $\sim 42$ per~cent of their initial orbital angular momentum, while their isotropic counterparts have lost only $\sim 19$ per~cent. The latter is consistent with a previous study of encounters between disc galaxy models dominated by isotropic dark haloes, which found average losses of $\sim 16$ per~cent \citep{Barnes2016}.

\subsubsection{Starting time}
\label{sec:starting_time}

As noted in \S~\ref{sec:Encounters}, most of our simulations start at $t_{0} = -200$ time units before pericenter, which places the two models relatively close. We therefore ran experiments starting at $t_{0} = -400$. These encounters used anisotropy radii $r_\mathrm{a} = 1.297$ and~$\infty$, and initial orbits with pericentric separations $r_\mathrm{p} = 1$,~$2$, and~$4$. To overcome run-to-run variation, we simulated $8$ realizations of each encounter (Table~\ref{tab:simulations}).

\begin{figure*}
    \centering
    \includegraphics[width=\textwidth]{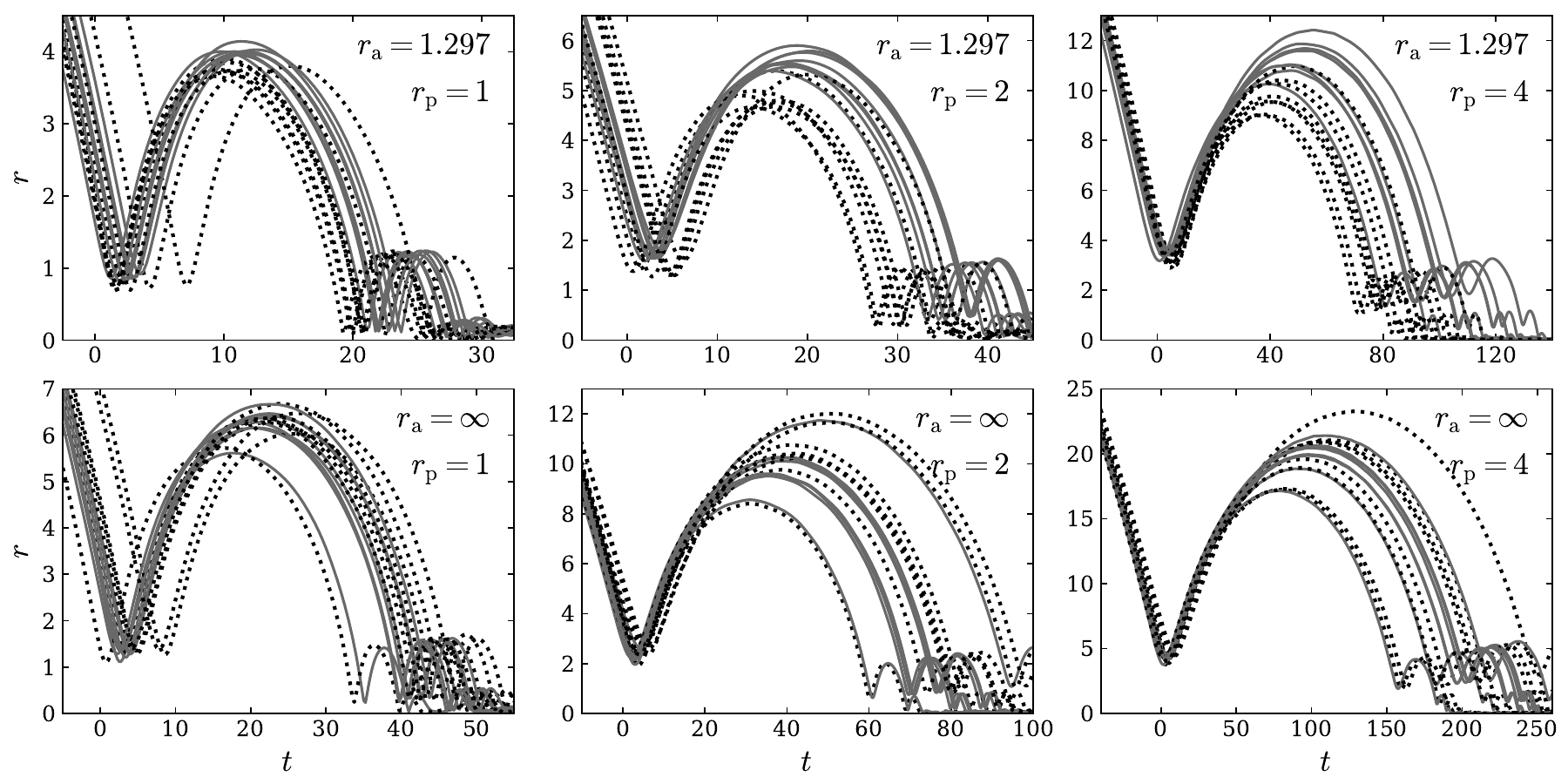}
    \caption{Effect of starting time on radial trajectories.  Runs starting at $t_{0} = -400$ (dotted) are compared with their counterparts starting at $t_{0} = -200$ (solid); all $8$ realizations of each ensemble are plotted.   Encounters of isotropic models (bottom) behave similarily for either choice of $t_{0}$. For anisotropic models (top), starting earlier lowers the apocentric separation and reduces the merger timescale.}
    \label{fig07}
\end{figure*}

We illustrate several consequences of changing $t_{0}$ in Fig.~\ref{fig07}. First, starting earlier appears to reduce the maximum separations that the anisotropic models (top panels) attain after first pericentre. For these models, starting earlier also appears to reduce the time $\Delta t$ between pericentres, although this trend is harder to disentangle due to run-to-run variations. In contrast, the isotropic models exhibit no obvious trends in separation or $\Delta t$ with starting time. Finally, starting earlier seems to increase the run to run variation; dotted curves, which show results for the earlier starting times, appear more spread out then their solid counterparts. This is further discussed in Appendix~A.

Fig.~\ref{fig08} illustrates the effect of starting times on pericentric parameters. Consistent with what we have just seen, anisotropic encounters which start earlier undergo more orbit decay, prior to first passage, than those which start later. The curves of constant angular momentum are clearly different in the anisotropic cases, but essentially indistinguishable in the isotropic cases. This difference arises because anisotropic systems begin transferring angular momentum earlier in the experiments starting at $t_{0}=-400$. For these experiments, up to $\sim 48$~per cent of the initial angular momentum is lost before first pericentre.

\begin{figure*}
    \centering
    \includegraphics[width=\textwidth]{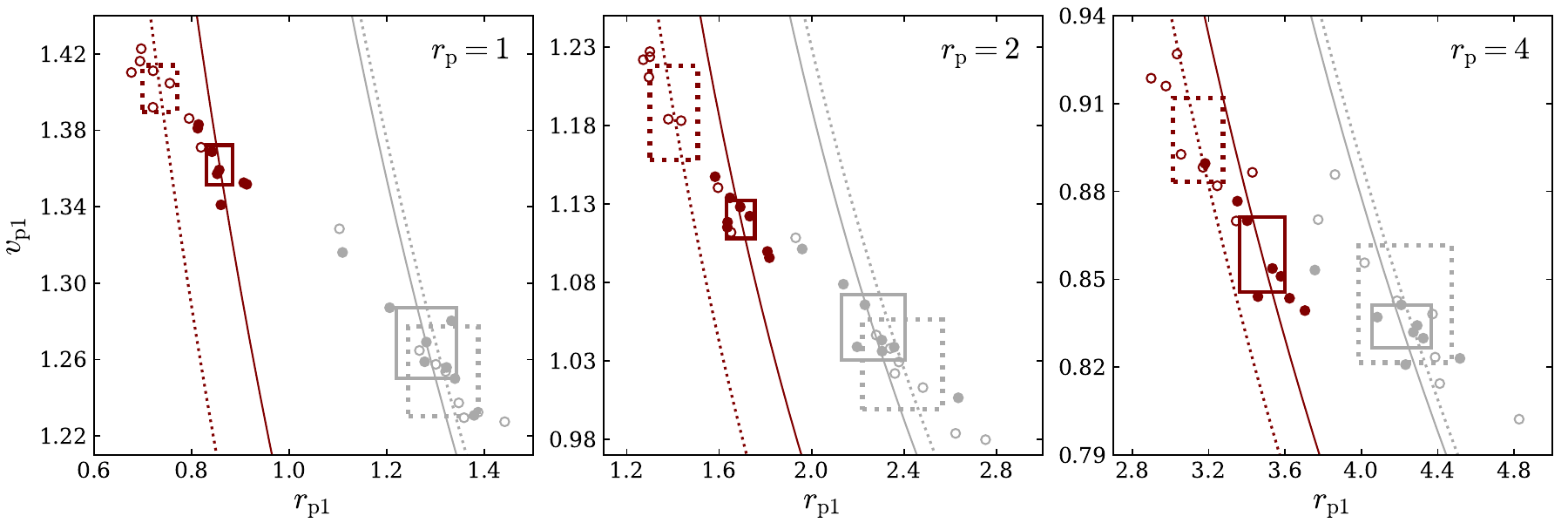}
    \caption{Effect of starting time on first passage parameters. Colors and symbols follow the convention used in Fig.~\ref{fig06}; in addition, open circles and dotted lines show results obtained by starting at $t_{0}=-400$. For the anisotropic models, earlier starting times produce closer and faster passages; while for the isotropic models the choice of starting time has no discernible effect.}
    \label{fig08}
\end{figure*}

\subsection{Response To Tidal Forces}
\label{sec:response_to_tidal_forces}

\begin{figure*}
   \centering
    \includegraphics[width=\textwidth]{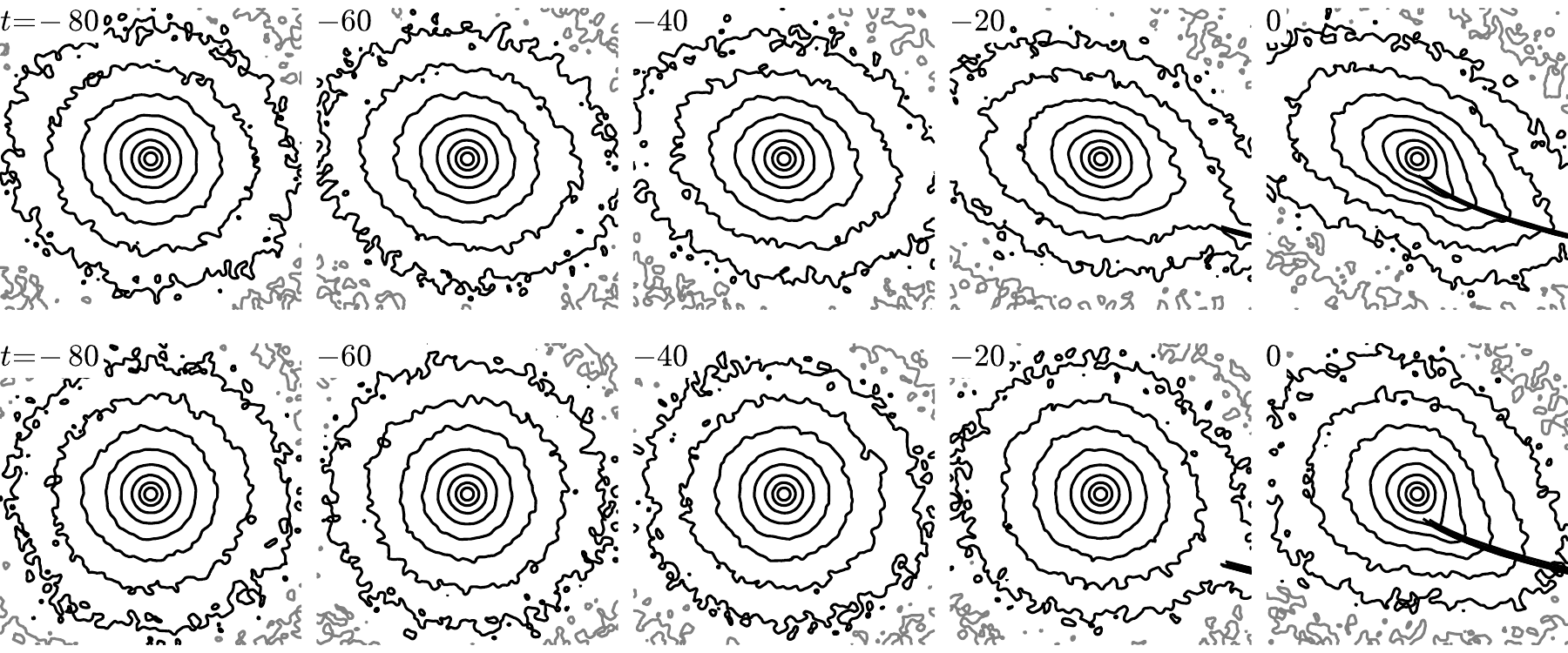}
    \caption{Response of one galaxy (contours) to the tidal field of its companion. This figure shows the same simulations presented in Fig.~\ref{fig03}, with anisotropic models above, and their isotropic counterparts below. Contours are spaced by factors of $\sqrt{8}$ in surface density; each frame is $32$ length units on a side. Black curves in the last two frames show trajectories of companion galaxies, which enter from the lower right.}
    \label{fig09}
\end{figure*}

Orbit decay in galaxy encounters is driven by the transfer of energy and angular momentum. If galaxies could not deform during tidal encounters, this transfer would be impossible. We therefore studied the shape of our galaxies prior to first pericentre, when they have not yet intermingled. Tidal deformation is already visible in Fig.~\ref{fig03}, but the response of each galaxy is hard to follow, as they partially overlap. By contouring the surface density for just one of the two galaxies, we can accurately visualize its response to the other galaxy's tidal field; Fig.~\ref{fig09} shows examples. The anisotropic simulations in the top row are conspicuously elongated, roughly towards the other galaxy, from $t=-60$ onward. The isotropic simulations in the bottom row show little sign of distortion until $t=-20$.

\begin{figure*}
    \centering
    \includegraphics[width=\textwidth]{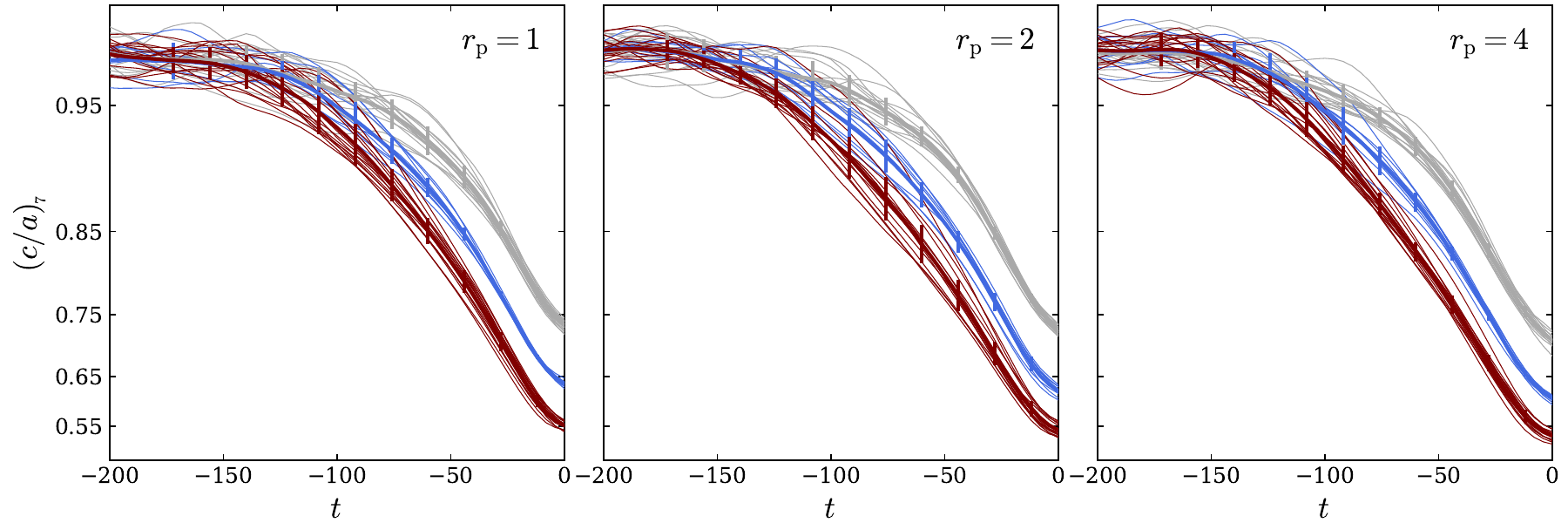}
    \caption{Effect of anisotropy on shapes of interacting models. Red, blue, and grey show results for $r_\mathrm{a}=1.297$, $r_\mathrm{a}=2.0$ and $r_\mathrm{a} = \infty$, respectively. Light curves are individual realizations, heavy lines represent sample means $\overline{(c/a)}$, and vertical bars show sample standard deviations $s_{(c/a)}$.  The vertical axis uses the transformation $y = - \log_{10}(1.1 - (c/a))$ to better show small departures from sphericity. Anisotropic models become distorted earlier and more strongly than their isotropic counterparts.}
    \label{fig10}
\end{figure*}

Fig.~\ref{fig10} compares shape responses of anisotropic and isotropic encounters for all $r_\mathrm{p}$ values. As in Sec.~\ref{sec:stability}, we section each realization by binding energy into $8$ shells; here we plot $(c/a)_{7}$, the minor-to-major axial ratio for the $7^\mathrm{th}$ shell. This probes the tidal response at the relatively large radii which drive orbit decay. As Fig.~\ref{fig09} already suggests, the anisotropic systems show a consistently earlier and stronger response to the other galaxy's tidal field. We observe a similar pattern of behavior for more tightly-bound shells, although the response is weaker since tides have less effect at smaller radii. 

\begin{figure*}
    \centering
    \includegraphics[width=\textwidth]{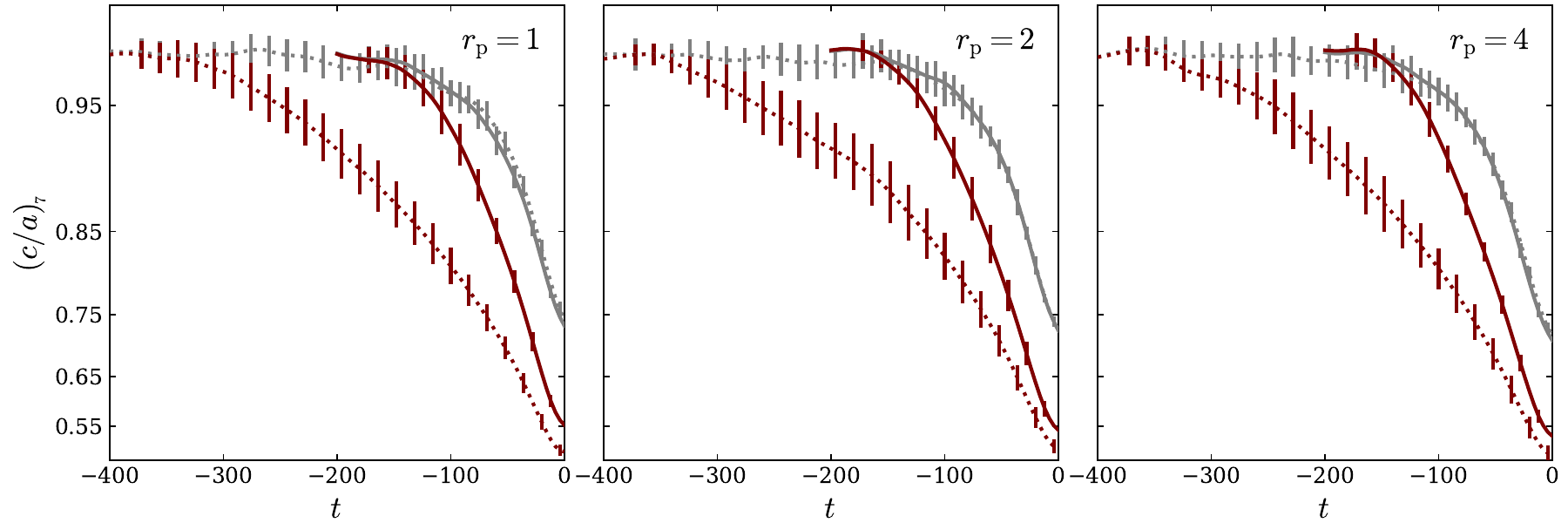}
    \caption{Effect of starting time on shape.  Dotted lines show simulations with $t_{0}=-400$, while solid lines show the matching simulations with $t_{0} = -200$. The vertical scaling and colors match Fig.~\ref{fig10}; to reduce clutter, individual runs are not plotted. Starting time has little effect on the shapes of isotropic models (grey); in contrast, starting earlier markedly increases the distortion of the anisotropic models (red).}
    \label{fig11}
\end{figure*}

Starting time has a clear effect on the orbital evolution of anisotropic models, and in Fig.~\ref{fig11} we show how starting time influences shape response. For isotropic models, the choice of starting time makes little difference; these systems remain nearly spherical until they draw close to each other. In contrast, starting time has a \textit{large} effect on the shapes of the anisotropic models; the encounters started at $t_{0} = -400$ are already exhibiting significant distortions (e.g., $(c/a)_{7} \simeq 0.9$) by $t_{0} = -200$.

Tidal distortion is necessary for orbit decay, but in order to actually transfer angular momentum, it's also necessary for this distortion to lag behind the current tidal forcing. This lag arises because a stellar system doesn't respond instantaneously to tidal perturbations; rather, a velocity perturbation imposed at time $t$ produces a density response at times $t^\prime > t$. In the case of a pair of galaxies making their initial approach, the major axis of each galaxy's tidal distortion does not point directly toward its companion, but rather toward where the companion \textit{was} at some earlier time. This misalignment produces a tidal torque which transfers orbital angular momentum to internal rotation.

\begin{figure*}
    \centering
    \includegraphics[width=\textwidth]{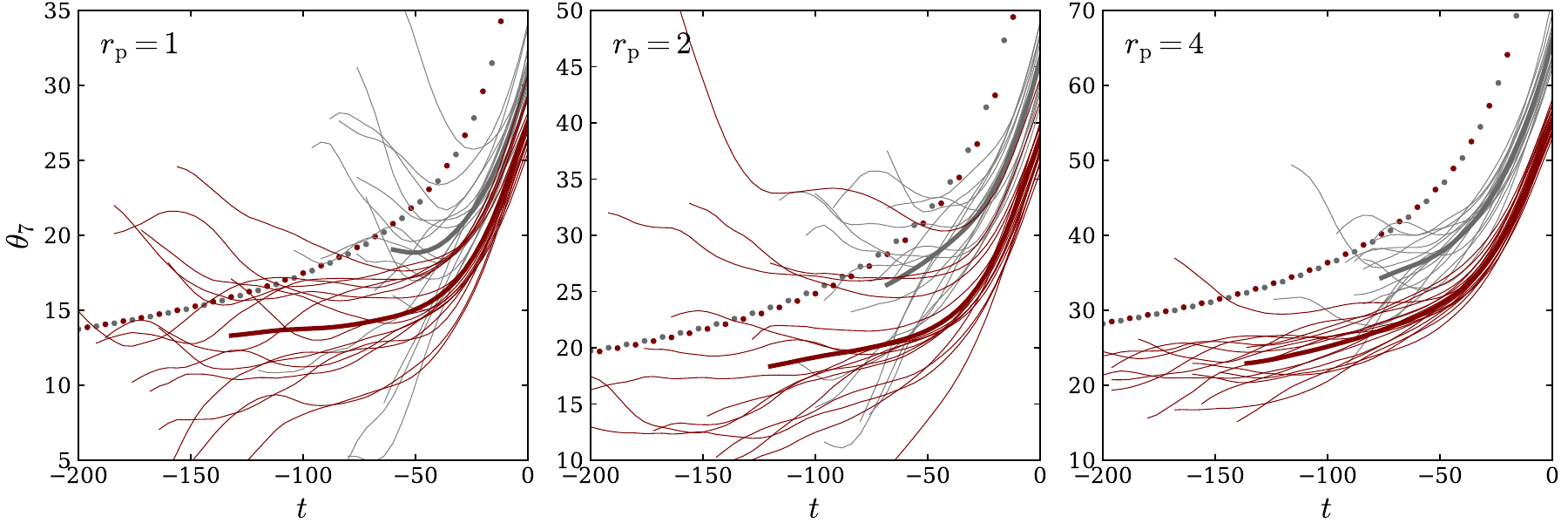}
    \caption{Orientation of tidal distortions vs.~time.  $\theta_{7}$ is the angle of the major axis of the $7^\mathrm{th}$ shell, measured clockwise from the $+x$ axis. Red and grey show results for $r_\mathrm{a} = 1.297$ and~$\infty$, respectively. Thin solid lines are individual realizations, while heavy solid lines are sample means. Small filled circles with alternating colors show the bearing of the companion galaxy.
    \label{fig12}}
\end{figure*}

Fig.~\ref{fig12} shows how orientations evolve during the approach to first pericentre.  We use the simulations starting at $t_{0} = -400$, which follow pre-encounter evolution more completely. Each galaxy is represented by a thin curve plotting the angle $\theta_7$ of its $7^\mathrm{th}$ shell, measured clockwise from the $+x$ axis; this curve is plotted from the time when the shell's axial ratio $(c/a)_{7}$ falls below $0.9$ ($0.95$) for the anisotropic (isotropic) models, respectively. The heavier solid curves are means computed over the entire sample. Finally the filled circles show angles to companion galaxies, again measured from the $+x$ axis.

This figure illustrates the delayed response necessary for orbit decay. It also points up another factor which plausibly accelerates orbit decay in encounters of anisotropic galaxies: the size of the lag. At $t \ll t_{0} = -400$ the companion galaxy's bearing would have been $\theta \simeq 0$ degrees (i.e., along the $+x$~axis), but by $t = t_{0}$ it has swung clockwise, reaching $\theta \simeq 11$~to $22$~degrees for $r_\mathrm{p} = 1$~to $4$, respectively. Until shortly before pericenter ($t = 0$), the distance between the galaxies scales approximately as $r \propto |t|^{2/3}$, and their mutual angular velocity scales as $\dot{\theta} \propto J r^{-2} \propto |t|^{-4/3}$. Anisotropic galaxies respond earlier to the tidal forcing (growing as $r^{-3}$), and their outer regions align roughly with the bearing of the other galaxy at \textit{early} times. In contrast, the isotropic galaxies don't respond until later, by which time the companion galaxy has swung further away from the $x$~axis. As a result, anisotropic galaxies lag more than their isotropic counterparts. This increased misalignment, along with the stronger tidal distortions already described, enable the anisotropic encounters to lose orbital angular momentum more rapidly.

\subsection{Perturbation Experiments}
\label{sec:perturbation_experiments}

To examine the effects of anisotropy in a less complicated setting, we simulated the response of single galaxy disturbed by a tide-like field. This allowed us to focus on the response without trying to disentangle the detailed interaction of two galaxies. We implemented the tide-like field by impulsively altering particle velocities. Inasmuch as \textit{any} perturbation may be approximated by a series of impulses, we expect that the response to a single impulse to be governed by the same physics which is relevant during the initial stages of a tidal encounter.

 Our first experiments adopted a linear tidal field aligned with the $x$-axis, mimicking the effect of a brief tidal perturbation due to a large mass located far away. However we found that this simple linear prescription had an oversized effect on particles at large radii. In particular, loosely bound particles far from the centre were given such large velocities that they became unbound.

To reduce the magnitude of the velocity perturbation on outlying particles, we next tried the following:
\begin{subequations}
\label{eq:perturbation}
\begin{align}
    &v_x'= v_x + \frac{2 A}{1+r/r_\mathrm{0}} \, x \, , \\
    &v_y'= v_y - \frac{A}{1+r/r_\mathrm{0}} \, y \, ,\\
    &v_z'= v_z - \frac{A}{1+r/r_\mathrm{0}} \, \, z \, .
\end{align}
\end{subequations}
The parameter $A$ controls the strength of the perturbation. We rather arbitrarily set $r_\mathrm{0}=1+\sqrt{2}$, which is the half mass radius of our original Hernquist model. In effect, this scheme follows the linear relationship at small radii but reduces the perturbation strength for $r > r_\mathrm{0}$. 

After perturbing the velocities, we simulated the model evolution and quantified the shape using the same methodology as in \S~\ref{sec:stability}. Experimenting with a range of values, we found that $A = 0.01$ produced a clear response which relaxed in a few hundred time units. The $7^\mathrm{th}$ shell gave the most unambiguous results, with more tightly bound shells behaving in a similar but less dramatic fashion.

\begin{figure}
    \centering
    \includegraphics[width=\columnwidth]{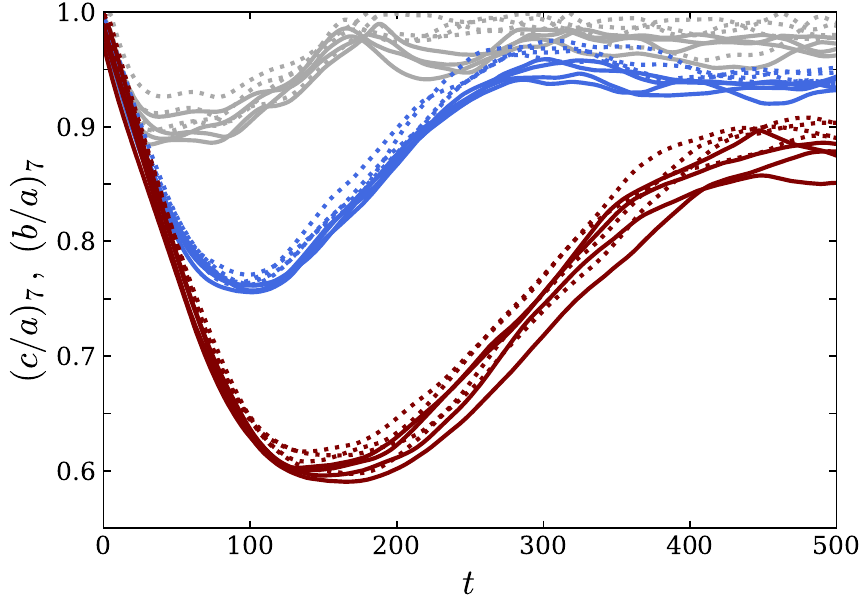}
    \caption{Shape response of galaxy models to tide-like perturbations. Red, blue, and grey curves show results for models with $r_\mathrm{a} = 1.297$, $r_\mathrm{a} = 2.0$ and $r_\mathrm{a} = \infty$. Solid and dotted curves show axial ratios $(c/a)_7$ and $(b/a)_7$, respectively. We used a perturbation strength of $A = 0.01$ (see equation~\ref{eq:perturbation}), and ran $4$ realizations for each model. The more anisotropic models show a stronger response to the imposed perturbation.}
    \label{fig13}
\end{figure}

Fig.~\ref{fig13} compares results for three different levels of anisotropy. Clearly, anisotropic models are more easily deformed than their isotropic counterparts. The $(b/a)_7$ ratios closely track the $(c/a)_7$ ratios, indicating these distortions are prolate, as expected given the symmetry of the imposed perturbation. These results may be compared with the results shown in Fig.~\ref{fig10}. Here the models are responding to a brief initial perturbation rather than the increasing tidal fields generated during initial approach, but in both cases the anisotropic models exhibit the strongest response. We argue that \textit{this stronger response accounts for the difference in orbit decay}.

\section{DISCUSSION}
\label{sec:discussion}

\begin{figure}
    \centering
    \includegraphics[width=60mm]{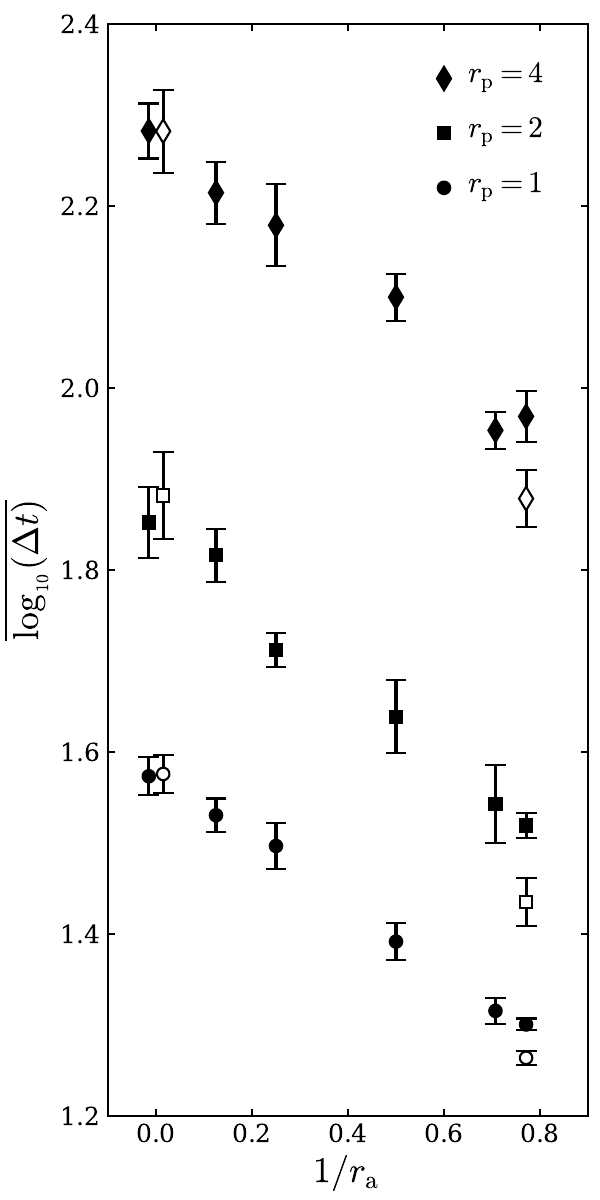}
    \caption{Relationship between orbit decay time $\Delta t$ and radial anisotropy, here plotted as $1/r_\mathrm{a}$ to include isotropic models. Markers show sample means $\overline{\log_{10}(\Delta t)}$, while vertical bars show $2 \sigma$ uncertainties. Solid and open markers are from runs starting at $t_{0} = -200$ and $-400$, respectively (plotted side by side where they would overlap). More anisotropic models undergo faster orbit decay.}
    \label{fig14}
\end{figure}

In encounters of equal-mass spherical systems, the rate of orbit decay is sensitive to radial velocity anisotropy. We demonstrated this by simulating encounters between idealized models with \citet{Hernquist1990} density profiles and \citet{Osipkov1979}-\citet{Merritt1985} velocity distributions. Fig.~\ref{fig14} summarizes results for all simulations listed in Table~\ref{tab:simulations}. The vertical axis is the log of $\Delta t$, the time between $1^\mathrm{st}$ and $2^\mathrm{nd}$ pericentric passages, averaged over all realizations in each ensemble. We plot the inverse of the anisotropy radius $r_\mathrm{a}$ on the horizontal axis to encompass both anisotropic and isotropic ($1/r_{\mathrm{a}} = 0$) models. Across the range of parabolic orbits in this study, the most anisotropic models merge in roughly half the time their isotropic counterparts require.

This result was prefigured by some earlier studies. \cite{AB2007} simulated the evolution of satellite galaxy systems with mass ratios of \hbox{$10 \!:\! 1$}, using primary galaxies with Osipkov-Merritt velocity distributions and \citet{Plummer1911} or \citet{Jaffe1983} density profiles. They found that anisotropy reduced the orbit decay time by $\sim 8$~per cent for the former, and by $\sim 22$~per cent for the latter. Their results are qualitatively consistent with, but smaller than, the effects we observe; the difference may be due to adopted mass ratio, emphasis on initially bound versus parabolic orbits, or degree of anisotropy.

Curiously, \citet{McMillan2007} did \textit{not} report any effect of halo aniso\-tropy on orbit decay in equal-mass disk galaxy mergers. As part of a multi-parameter study of remnant structure, they ran experiments with both radially and tangentially anisotropic haloes. They found that initial anisotropy has a modest effect on remnant halo density profiles \citep[consistent with][]{White1979}, but very little influence on other remnant properties. We estimate that their models were only \textit{mildly} anisotropic, perhaps comparable to our models with anisotropy radius $r_\mathrm{a} = 4$, and conjecture that the resulting decrease of orbit decay times was not dramatic enough to be mentioned in a study focused on remnant haloes.

More recently, \cite{Vasiliev2022} simulated orbital evolution in unequal-mass systems to study radialization of satellite orbits. They present two simulations involving anisotropic primary galaxies, with primary-to-satellite mass ratios of $10 \!:\! 1$, and find that radial (tangential) anisotropy enhances (suppresses) radialization. Close inspection of their Fig.~3 also suggests that their radially anisotropic model may have merged more rapidly than its isotropic counterpart.

The immediate reason for faster mergers between anisotropic systems appears straightforward: anisotropic models are more easily deformed during tidal interactions. We have shown that anisotropic models become more non-spherical than their isotropic counterparts during the initial stages of an encounter; it's likely that this earlier response also explains their larger misalignments with the present positions of their companions (\S~\ref{sec:response_to_tidal_forces}). This accelerates the transfer of orbital angular momentum to internal degrees of freedom, speeding up the merging process. Since our interpretation is based on very general considerations, it's likely to apply in any tidal encounter between galaxies with significant radial anisotropy.

Simulations of isolated models support the idea that radial anisotropy makes systems more sensitive to non-radial perturbations (\S~\ref{sec:perturbation_experiments}). Several previous studies have also found that radial anisotropy increases galactic responses to tides. In simulated Milky~Way-LMC encounters, \cite{Rozier2022} report that the wake induced in the Milky~Way's \textit{stellar} halo by the passage of the LMC is larger if the halo is radially anisotropic, and smaller if it is tangentially anisotropic (in both cases the dark halo was isotropic). This suggests the strength of the response does not depend on self-gravity, since the stellar halo's mass is negligible.  Similarly, \cite{Vasiliev2024} simulated Milky~Way-LMC encounters and found that the wake induced by the LMC was larger when the Milky~Way's dark halo was radially anisotropic.

The reason for this larger response is not immediately obvious. It’s intuitively appealing to think that the anisotropic pressure in our models explains why these systems are more responsive to non-radial perturbations. But \citet{PP1987}, in a closely related context, showed that pressure anisotropy \textit{fails} to predict the onset of the radial orbit instability. They present a physical interpretation of the instability based on the precession rates of very eccentric orbits subject to a bar-like perturbation; we suggest that a similar mechanism may be at work here. Linear theory calculations \citep[e.g.,][Ch.~5]{BT2008} and test-particle simulations may help clarify the basis for the stronger response of anisotropic systems.

We found that the choice of starting time has a significant influence on encounters of anisotropic systems. In Fig.~\ref{fig14}, the open markers represent runs starting at $t_{0} = -400$ instead of the default $t_{0} = -200$. The choice of starting time appears to have no effect on the evolution of the isotropic models; for such systems, the potential-energy criterion $\delta U \ll G M^2/r_{0}$ (see equation~\ref{eq:deltaU}) appears adequate. However, the anisotropic models behave differently; those started earlier merge faster. This supports the idea that anisotropic galaxies are susceptible to tidal deformations quite early in the encounter process. It is possible that even $t_{0} = -400$ is not early enough to fully capture the effect of tidal interactions between these anisotropic systems. The importance of early tidal interaction was not appreciated in previous studies; computational considerations generally favor launching galaxies in relatively close proximity. Given that the choice of initial separation appears to have significant consequences for encounters involving anisotropic systems, it should be explicitly discussed in future studies.

Our results are relevant to `idealised' simulations in which two or more equilibrium galaxy models are allowed to interact with each other. It seems plausible that haloes formed partly by gravitational collapse would have predominantly radial orbits. Idealised models could be made more realistic by including some level of radial anisotropy. Below, we outline three possible applications.

First, shorter merger time scales due to halo anisotropy may influence estimates of the merger rate. \citet{Lotz}~used merger time scales derived from idealised simulations, together with observational counts of interacting galaxies, to infer rates of galaxy mergers. The haloes used in their simulations were very close to isotropic, potentially overestimating the merger time scale. If galaxies take less time to merge, then the merger rate must be higher to account for the number of merging galaxies observed.

Second, faster orbit decay may help explain the persistence of tidal tails in merger remnants. Tails formed during interactions of disc galaxies embedded in massive dark haloes generally remain bound and may fall back before the galaxies finally merge \citep[e.g.,][and references therein]{Barnes2016}. Such complete recapture is incompatible with observations of long-tailed merger remnants such as NGC~7252 \citep{Schweizer1982}. Faster orbit decay in systems with anisotropic haloes may make it easier for galaxies to merge before their tails fall back. We hope to explore this question in future work.

Third, semi-analytic models of galaxy formation often estimate merger times from simple formulas based on the \cite{Chandrasekhar1943} treatment of dynamical friction. \citet*{BKMQ2008} use simulated mergers of isotropic haloes to argue that these simple formulas underestimate merging times by factors of $\sim 1.7$ or more, distorting key semi-analytic predictions such as the build-up of central galaxy stellar masses. Including anisotropy in the simulations would presumably -- for \textit{purely} fortuitous reasons -- have reduced the discrepancy with the dynamical friction formulas.

Possible directions for future work include:
\begin{enumerate}

    \item Encounter dynamics. The present study could be extended to head-on encounters (in which angular momentum transfer does not play a role), and to encounters involving tangentially anisotropic models. The early stages of equal-mass tidal encounters may be amenable to linear theory calculations.
    
    \item Unequal-mass interactions. We expect that our results also apply to encounters between systems with different masses. While \citet{AB2007} found that satellite galaxies orbiting in radially aniso\-tropic haloes suffer more rapid orbit decay, it would be worthwhile to understand why the effects they report are smaller than those we observe.
    
    \item Remnant structure. \citet{McMillan2007} found that moderate levels of initial anisotropy have little effect on remnant anisotropy, and only slight effects on remnant shape. However, we speculate that the earlier transfer of angular momentum in encounters between anisotropic systems may produce remnants with more of their angular momentum retained at large radii.
    
\end{enumerate}

The implications of our results for cosmological simulations, or for real galaxies, are somewhat more complex. Simulations in which structure grows from linear perturbations automatically incorporate halo anisotropy. In the specific case of $\Lambda$CDM, \citet{Wojtak+2005} found that haloes in dark matter only simulations are approximately isotropic at their centres but radially anisotropic further out, rising to $\beta \simeq 0.5$ at the virial radius. Such haloes, while less anisotropic than our most extreme examples, may be anisotropic enough to influence merger times. However, haloes in cosmological simulations don't form as spherical systems; instead, they are generically triaxial, and typically aligned with their neighbors and with the filamentary pattern of large-scale structure \citep[e.g.,][]{BS2005,V+2018,D+2023}. Idealised simulations don't capture these phenomena. But the mutual alignments created when idealised anisotropic models interact qualitatively mimic the long-range alignments seen in cosmological simulations; the subsequent evolution of such models may be a useful analog to the behavior of real systems.

\subsection{Summary}
\label{Summary}

Our investigation of radial anisotropy in encounters of spherical galaxy models has shown that anisotropic models merge faster than their isotropic counterparts. In comparison to isotropic models, anisotropic models are more susceptible to tidal deformation. This accelerates the process of momentum transfer and orbit decay. Measurements of deformation confirm that anisotropic models become more elongated than their isotropic counterparts. During encounters, anisotropic models arrive at first passage having already lost significant amounts of angular momentum.

\section*{ACKNOWLEDGEMENTS}

We thank the referee for a constructive and stimulating report, and Colby Haggerty and Jennifer Lotz for helpful discussions. JEB is grateful to the Yukawa Institute for Theoretical Physics, Kyoto University, for generous hospitality in Fall 2022 and Summer 2023, and especially to Atsushi Taruya for sponsoring me during these visits. LS thanks the members of the Spring 2022 Senior Research Project class for feedback and suggestions.

\section*{DATA AVABILITY}

Simulation data and software are available by request to barnes@hawaii.edu.

\appendix

\section{RUN-TO-RUN VARIATIONS}

We have taken advantage of the relatively low computational cost of our numerical experiments by running from $4$ to $8$ independent realizations of each physical scenario (Table~\ref{tab:simulations}). Run-to-run variations among ostensibly equivalent realizations can be surprisingly large; in some ensembles, the spread in the orbit decay time $\Delta t$ is $\sim 10$ per~cent of the mean value. Below, we show that these variations are primarily driven by the run-to-run spread in orbital angular momentum.

A galaxy realization is generated by independently drawing $N$ particle positions $\mathbf{r}_i$ and velocities $\mathbf{v}_i$ from a DF (\S~\ref{sec:nbody}). Integrated quantities therefore have variations of $O(N^{-1/2})$ from one realization to the next. For example, consider the $x$-component of the centre-of-mass position, which is $\overline{x} \equiv N^{-1} \sum_{i=1}^{N} x_i$ for $N$ equal-mass particles. In the large-$N$ limit, $\overline{x}$ has a normal distribution, with a mean of $0$ and a standard deviation of $\sigma_{\overline{x}} = x_\mathrm{rms} N^{-1/2}$, where $x_\mathrm{rms} \equiv (N^{-1} \sum_{i=1}^{N} x_i^2)^{1/2}$ is the RMS value of the $x$ coordinate. Since the models are spherically symmetric, the $y$~and $z$-components have the same distribution. Similar reasoning applies to the centre-of-mass velocity, and to the net angular momentum. The net binding energy $\mathcal{E}$ also varies about its mean value $\mathcal{E} \simeq -0.0850$; numerical tests confirm that $\sigma_\mathcal{E} \propto N^{-1/2}$. Table~\ref{tab:model_variation} lists dispersions in position, velocity, angular momentum, and energy for isotropic models realized with $N = 65536$ particles.

\begin{table}
    \centering
    \caption{Standard deviations in 1-D centre-of-mass position, velocity, angular momentum, and in net binding energy for realizations with $N = 65536$ particles.}
    \begin{tabular}{lll}
    \noalign{\bigskip} \hline \noalign{\smallskip}
    \textbf{Quantity} & \textbf{Symbol} & \textbf{Value} \\
    \noalign{\smallskip} \hline \noalign{\smallskip}
    C.M.~Position    & $\sigma_{r}$ & 0.0464 \\
    C.M.~Velocity    & $\sigma_{v}$ & 0.000938 \\
    Angular Momentum & $\sigma_\mathcal{J}$ & 0.00376 \\
    Binding Energy   & $\sigma_\mathcal{E}$ & 0.000792 \\
    \noalign{\smallskip} \hline
    \end{tabular}
    \label{tab:model_variation}
\end{table}

To set up an encounter, two such galaxy realizations are translated by $\pm \frac{1}{2} \mathbf{r}_{0}$ in position and $\pm \frac{1}{2} \mathbf{v}_{0}$ in velocity, as specified by equation~(\ref{eq:sys_dist_func}). The initial parabolic orbit has angular momentum about the $z$-axis
\begin{equation}
    \mathcal{J}_\mathrm{orb} = \sqrt{G M^3 r_\mathrm{p}} = \frac{1}{2} M \, (\mathbf{r}_{0} \mathbf{\times} \mathbf{v}_{0}) \,\mathbf{\cdot}\, \hat{\mathbf{z}} \, ,
\end{equation}
where $M$ is the mass of a single model, and $\hat{\mathbf{z}}$ is a unit vector along the $z$-axis. For finite $N$, however, the initial centre-of-mass position $\mathbf{r}_{\ell}$ and velocity $\mathbf{v}_{\ell}$ of galaxy realization $\ell = 1, \, 2$ will be offset by random vectors of magnitude $\sim \sigma_{r}$ and $\sim \sigma_{v}$, respectively. As a result, the effective orbital angular momentum
\begin{equation}
    \mathcal{J}_\mathrm{eff} = \sum_{\ell=1}^2 M \, (\mathbf{r}_{\ell} \mathbf{\times} \mathbf{v}_{\ell}) \,\mathbf{\cdot}\, \hat{\mathbf{z}} \, ,
\end{equation}
includes stochastic terms proportional to $\sigma_{r}$ and $\sigma_{v}$.

For the encounters considered here, the run-to-run variation in $\mathcal{J}_\mathrm{eff}$ is dominated by the variation in centre-of-mass velocity. This is a direct consequence of our efforts to keep the models from overlapping too much at the start of the simulations. As an example, starting the $r_\mathrm{p} = 1$ orbit at time $t_{0} = -200$ initially places the models $r_{0} \simeq 70.2$ length units apart. The perpendicular orbital velocity of each model is then only $\sim 15$ times larger than the variation in centre-of-mass velocity for $N = 2^{16}$ particles. In brief, it's hard to come close to the intended angular momentum when galaxy realizations must be launched from far apart.

More formally, the variance of $\mathcal{J}_\mathrm{eff}$ is
\begin{equation}
    \mathrm{Var}(\mathcal{J}_\mathrm{eff}) =
        2 \, ( \texthalf r_{0} \sigma_{v})^2 +
        2 \, ( \texthalf v_{0} \sigma_{r})^2 \, .      
\end{equation}
The first term in this sum accounts for $\sim 97$ per~cent of the variance for orbits starting at $t_{0} = -200$, and $\sim 99$ per~cent for orbits starting at $t_{0} = -400$. For the various choices of $r_\mathrm{p}$ and $t_{0}$ used in this work, the standard deviation of $\mathcal{J}_\mathrm{eff}$ is between $2.2$ per~cent and $7.2$ per~cent of its mean value.

In passing, it's worth noting that run-to-run variations in orbital binding energy $\mathcal{E}_\mathrm{eff} = \frac{1}{2} M (v_{1}^2 + v_{2}^2) - G M^2 / | \mathbf{r}_{1} - \mathbf{r}_{2} |$ are much smaller. For all intents and purposes, the orbits used in our simulations are mono-energetic.

\begin{figure}
    \centering
    \includegraphics[width=\columnwidth]{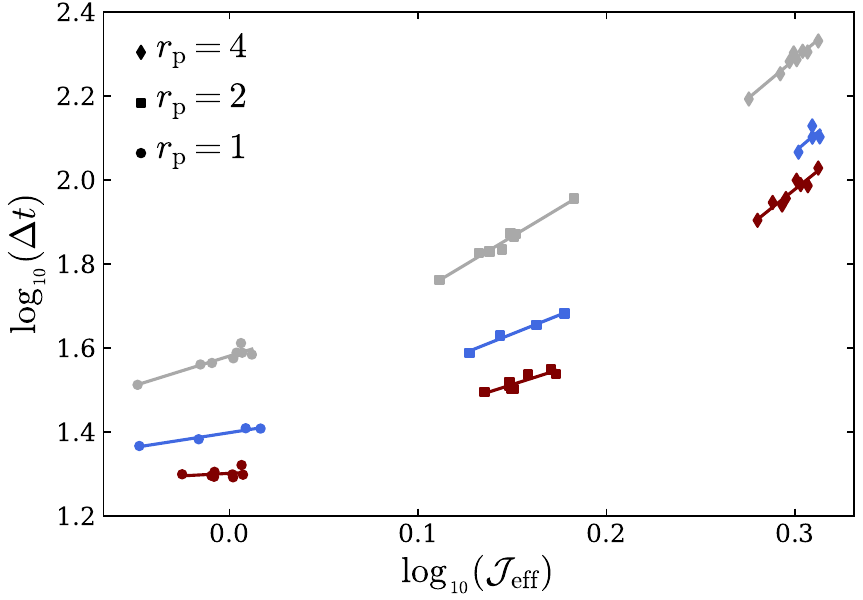}
    \caption{Relationship between orbit decay time $\Delta t$ and effective angular momentum $\mathcal{J}_\mathrm{eff}$.  Results are plotted for runs starting at $t_{0} = -200$ with anisotropy radii $r_\mathrm{a} = 1.297$, $2$, and $\infty$ (red, blue, and grey, respectively).  Points are individual runs; lines are local linear fits.}
    \label{fig15}
\end{figure}

Other things being equal, systems started with more orbital angular momentum should take longer to merge. Figs.~\ref{fig05} and~\ref{fig14} suggest that $\Delta t$, the time between first and second pericentric passages, is an increasing function of pericentric separation $r_\mathrm{p}$, which in turn is $\propto \mathcal{J}_\mathrm{orb}^2$. Fig.~\ref{fig15} provides further support. For each given $r_\mathrm{a}$, there is clearly a well-defined relationship between $\Delta t$ and $\mathcal{J}_\mathrm{eff}$. This figure also shows that, within each ensemble, \textit{run-to-run variation in $\mathcal{J}_\mathrm{eff}$ accounts for most of the variation in $\Delta t$.}

\begin{figure*}
    \centering
    \includegraphics[width=0.75\textwidth]{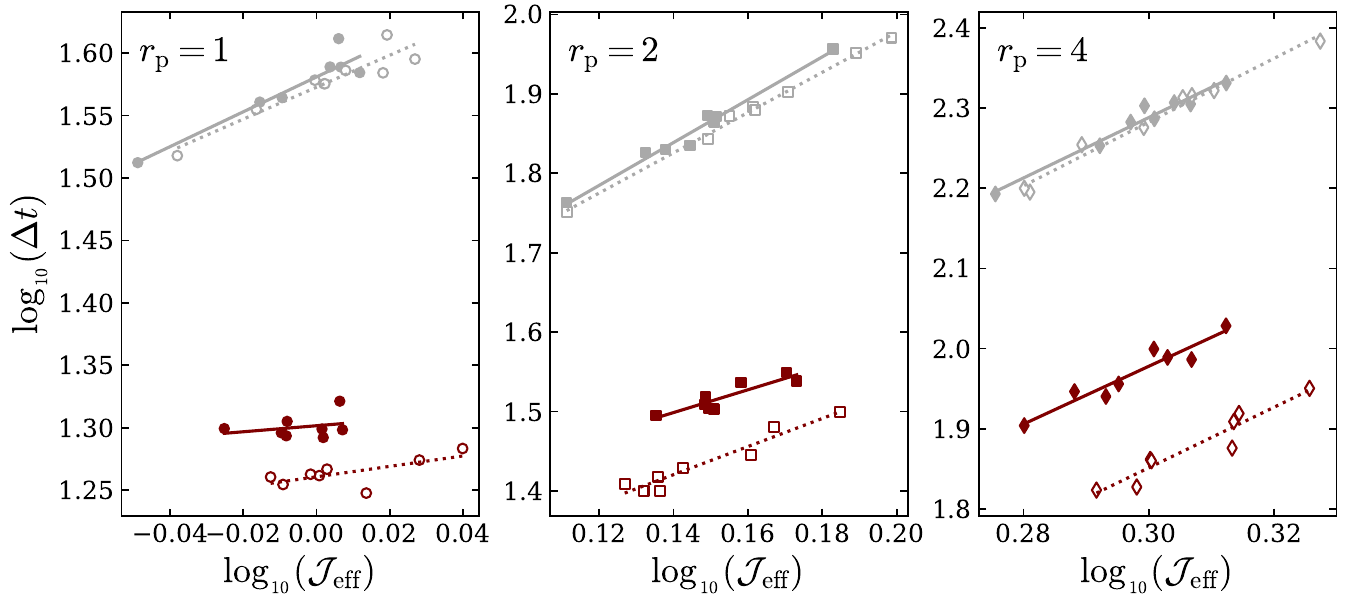}
    \caption{Effect of starting time on the relationship between $\Delta t$ and $\mathcal{J}_\mathrm{eff}$. Filled symbols and solid lines represent $t_{0} = -200$ runs; open symbols and dotted lines represent $t_{0} = -400$ runs.}
    \label{fig16}
\end{figure*}

The sensitivity of anisotropic mergers to the choice of starting time $t_{0}$ becomes very clear when results are plotted on the $\Delta t$ vs.~$\mathcal{J}_\mathrm{eff}$ plane. Fig.~\ref{fig16} compares $t_{0} = -400$ runs (open symbols, dotted lines) with their $t_{0} = -200$ counterparts (filled symbols, solid lines). This plot confirms that isotropic encounters (grey) merge in the same time for either choice of $t_{0}$, while anisotropic encounters (red) merge faster when started earlier. 

The correlation between $\Delta t$ and $\mathcal{J}_\mathrm{eff}$ may be useful to reduce uncertainties. For example, when averaging over the members of an ensemble, more weight could be given to those runs with $\mathcal{J}_\mathrm{eff}$ values closer to the ideal $\mathcal{J}_\mathrm{orb}$. An even better strategy might be to use linear fits like those plotted in Fig.~\ref{fig15} to estimate what $\Delta t$ would result from an encounter with $\mathcal{J}_\mathrm{eff} = \mathcal{J}_\mathrm{orb}$. The uncertainty in this estimate could be inferred from the residuals of the linear relationship.

Another option, of course, is to subtract the centre-of-mass position and velocity from each galaxy realization before it is translated to its launch coordinates. This strategy ensures that $\mathcal{J}_\mathrm{eff}$ for each encounter is exactly equal to $\mathcal{J}_\mathrm{orb}$. The drawback, in our view, is that it introduces approximations into equation~(\ref{eq:sys_dist_func}). In effect, subtracting centre-of-mass coordinates simply `papers over' one obvious manifestation of Monte-Carlo sampling.

Finally, we note that simulations starting earlier that $t_{0} = -400$ would suffer from even more dramatic run-to-run variations. It may be interesting to try earlier starting times, since we are not sure that our experiments start early enough to capture the full effect of anisotropy on orbit decay. Large ensembles or larger values of $N$ might be needed to overcome uncertainties in such experiments.

\bsp	
\label{lastpage}


\begin{thebibliography}{}

    \raggedright

    \bibitem[Amorisco(2017)]{Amorisco2017} Amorisco, N.~C.\ 2017, \mnras, 464, 2882 
    
    \bibitem[Antonov(1973)]{Antonov1973} Antonov, V.~A., 1973, \sovast, 17, 428

    \bibitem[Arena \& Bertin(2007)]{AB2007} Arena, S.~E. \& Bertin, G.\ 2007, \aap, 463, 921 

    \bibitem[Bailin \& Steinmetz(2005)]{BS2005} Bailin, J., Steinmetz, M.\ 2005, \apj, 627, 647
    
    \bibitem[Barnes(2012)]{Barnes2012} Barnes, J.~E., 2012, \mnras, 425, 1104 

    \bibitem[Barnes(2016)]{Barnes2016} Barnes, J.~E., 2016, \mnras, 455, 1957 
    
    \bibitem[Barnes \& Hut(1986)]{BH1986} Barnes, J., Hut, P., 1986, \nat, 324, 446 
    
    \bibitem[Barnes et al.(1986)]{BGH1986} Barnes, J., Goodman, J., Hut, P., 1986, \apj, 300, 112 
        
    \bibitem[Binney \& Tremaine(2008)]{BT2008} Binney, J., Tremaine, S., 2008, Galactic Dynamics: Second Edition. Princeton University Press, Princeton, NJ

    \bibitem[Boylan-Kolchin, Ma, \& Quataert(2008)]{BKMQ2008} Boylan-Kolchin, M., Ma, C.-P., Quataert, E.\ 2008, \mnras, 383, 93 
    
    \bibitem[Buyle et al.(2007)]{Buyle2007} Buyle, P., van Hese, E., de Rijcke, S., Dejonghe, H., 2007, \mnras, 375, 1157 

    \bibitem[Chandrasekhar(1943)]{Chandrasekhar1943} Chandrasekhar, S.\ 1943, \apj, 97, 255. 
    
    \bibitem[Cox et al.(2006)]{Cox2006} Cox, T.~J., Jonsson, P., Primack, J.~R., Somerville, R.~S., 2006, \mnras, 373, 1013 

    \bibitem[Debattista \& Sellwood(2000)]{DS2000} Debattista, V.~P., Sellwood, J.~A., 2000, \apj, 543, 704

    \bibitem[Delgrado et al.(2023)]{D+2023} Delgrado, A.~M. et al.\ 2023, \mnras, 523, 5899

    \bibitem[Drakos et al.(2019)]{Drakos2019} Drakos, N.~E., Taylor, J.~E., Berrouet, A., Robotham, A.~S.~G., Power, C., 2019, \mnras, 487, 993

    \bibitem[Fulton \& Barnes(2001)]{FB2001} Fulton, E., Barnes, J.~E., 2001, Ap\&SS, 276, 851
    
    \bibitem[Hernquist(1990)]{Hernquist1990} Hernquist, L., 1990, \apj, 356, 359 

    \bibitem[Jaffe(1983)]{Jaffe1983} Jaffe, W., 1983, \mnras, 202, 995
    
    \bibitem[Lotz et al.(2008)]{Lotz} Lotz, J.~M., Jonsson, P., Cox, T.~J., Primack, J., 2008, \mnras, 391, 1137 
    
    \bibitem[Makino \& Hut(1997)]{MH1997} Makino, J., Hut, P., 1997, \apj, 481, 83 

    \bibitem[McMillan et al.(2007)]{McMillan2007} McMillan, P.~J., Athanassoula, E., \& Dehnen, W.\ 2007, \mnras, 376, 1261. 
    
    \bibitem[Merritt(1985)]{Merritt1985} Merritt, D., 1985, \aj, 90, 1027 

    \bibitem[Merritt \& Aguilar(1985)]{MA1985} Merritt, D., Aguilar, L.~A., 1985, \mnras, 217, 787
    
    \bibitem[Meza \& Zamorano(1997)]{MZ1997} Meza, A., Zamorano, N., 1997, \apj, 490, 136. 
    
    \bibitem[Osipkov(1979)]{Osipkov1979} Osipkov, L.~P., 1979, Azh, 5, 77L

    \bibitem[Palmer \& Papaloizou(1987)]{PP1987} Palmer \& Papaloizou, 1987, MNRAS 224, 1043

    \bibitem[Plummer(1911)]{Plummer1911} Plummer, H.~C., 1911, \mnras, 71, 460
 
    \bibitem[Rozier et al.(2022)]{Rozier2022} Rozier, S., Famaey, B., Siebert, A., et al.\ 2022, \apj, 933, 113 
    
    \bibitem[Schweizer(1982)]{Schweizer1982} Schweizer, F., 1982, \apj, 252, 455. 
    
    \bibitem[Springel \& White(1999)]{SW1999} Springel, V., White, S.~D.~M., 1999, \mnras, 307, 162. 
    
    \bibitem[Toomre \& Toomre(1972)]{TT1972} Toomre, A., Toomre, J., 1972, \apj, 178, 623. 
    
    \bibitem[van Albada \& van Gorkom(1977)]{VV1977} van Albada, T.~S., van Gorkom, J.~H., 1977, \aap, 54, 121

    \bibitem[Vasiliev et al.(2022)]{Vasiliev2022} Vasiliev, E., Belokurov, V., \& Evans, N.~W.\ 2022, \apj, 926, 203 

    \bibitem[Vasiliev(2024)]{Vasiliev2024} Vasiliev, E.\ 2024, \mnras, 527, 437. 

    \bibitem[Veena et al.(2018)]{V+2018} Veena, G.~P., Cautun, M., van de Weygaert, R., Tempel, E., Jones, B.~J.~T., Rieder, S., Frenk, C.~S., 2018, \mnras, 481, 414
    
    \bibitem[Vergne \& Muzzio(1995)]{VM1995} Vergne, M.~M., Muzzio, J.~C., 1995, \mnras, 276, 439 
    
    \bibitem[Villumsen(1982)]{V1982} Villumsen, J.~V., 1982, \mnras, 199, 493 
    
    \bibitem[Villumsen(1983)]{V1983} Villumsen, J.~V., 1983, \mnras, 204, 219 
    
    \bibitem[White(1978)]{White1978} White, S.~D.~M., 1978, \mnras, 184, 185 
    
    \bibitem[White(1979)]{White1979} White, S.~D.~M., 1979, \mnras, 189, 831 
    
    \bibitem[Wojtak et al.(2005)]{Wojtak+2005} Wojtak, R., {\L}okas, E.~L., Gottl{\"o}ber, S., Mamon, G.~A., 2005, \mnras, 361, L1 

\end{thebibliography}
\end{document}